\documentclass[12pt,thmsa]{article}%
\usepackage{amsmath}
\usepackage{amssymb}
\usepackage{sw20jart}%
\usepackage{amsfonts}%
\usepackage{graphicx}
\providecommand{\U}[1]{\protect\rule{.1in}{.1in}}
\begin{document}

\title{AMBIGUOUS VOLATILITY, POSSIBILITY AND UTILITY IN CONTINUOUS
TIME\thanks{Department of Economics, Boston University, lepstein@bu.edu and
School of Mathematics, Shandong University, jsl@sdu.edu.cn. We gratefully
acknowledge the financial support of the National Science Foundation (awards
SES-0917740 and 1216339), the National Basic Research Program of China
(Program 973, award 2007CB814901) and the National Natural Science Foundation
of China (award 10871118). We have benefited also from discussions with Shige
Peng, Zengjing Chen, Mingshang Hu, Jin Ma, Guihai Zhao and especially Jianfeng
Zhang. Epstein is grateful also for the generous hospitality of CIRANO where
some of this work was completed during an extended visit. The original version
of this paper, first posted March 5, 2011, contained applications to asset
pricing which are now contained in Epstein and Ji \cite{ej}. Marcel Nutz
pointed out an error in a previous version.}}
\author{Larry G. Epstein
\and Shaolin Ji}
\maketitle
\date{}

\begin{abstract}
This paper formulates a model of utility for a continuous time framework that
captures the decision-maker's concern with ambiguity about both the drift and
volatility of the driving process. At a technical level, the analysis requires
a significant departure from existing continuous time modeling because it
cannot be done within a probability space framework. This is because ambiguity
about volatility leads invariably to a set of nonequivalent priors, that is,
to priors that disagree about which scenarios are possible.

\medskip

\textit{Key words}: ambiguity, recursive utility, G-Brownian motion,
undominated measures, quasisure analysis, robust stochastic volatility

\end{abstract}

\newpage

\section{Introduction}

This paper formulates a model of utility for a continuous time framework that
captures the decision-maker's concern with ambiguity or model uncertainty. The
paper's novelty lies in the range of model uncertainty that is accommodated.
Specifically, aversion to ambiguity about both drift and volatility is
captured. At a technical level, the analysis requires a significant departure
from existing continuous time modeling because it cannot be done within a
probability space framework. This is because ambiguity about volatility leads
invariably to an undominated set of priors. In fact, priors are typically
nonequivalent (not mutually absolutely continuous) - they disagree about which
scenarios are possible.

The model of utility is a continuous time version of multiple priors (or
maxmin) utility formulated by Gilboa and Schmeidler \cite{gs} for a static
setting. Related continuous time models are provided by Chen and Epstein
\cite{CE} and also Hansen, Sargent and coauthors (see Anderson et al.
\cite{ahs}, for example).\footnote{The discrete time counterpart of the former
is axiomatized in Epstein and Schneider \cite{es2003}.} In these papers,
ambiguity is modeled so as to retain the property that all priors are
equivalent. This universal restriction is driven by the technical demands of
continuous time modeling, specifically by the need to work within a
probability space framework. Notably, in order to describe ambiguity authors
invariably rely on Girsanov's theorem for changing measures. It provides a
tractable characterization of alternative hypotheses about the true
probability law, but it also limits alternative hypotheses to correspond to
measures that are both mutually equivalent and that differ from one another
\emph{only} in what they imply about drift. This paper defines a more general
framework within which one can model the utility of an individual who is not
completely confident in any single probability law for volatility or in which
future events are possible.

Economic motivation for our model, some applications to asset pricing, and
also an informal intuitive outline of the construction of utility (without
proofs), are provided in a companion paper Epstein and Ji \cite{ej}%
.\footnote{Economic motivation is provided in part by the importance of
stochastic volatility modeling in both financial economics and macroeconomics,
the evidence that the dynamics of volatility are complicated and difficult to
pin down empirically, and the presumption that complete confidence in any
single parametric specification is often unwarranted and implausible.} The
present paper provides a mathematically rigorous treatment of utility. Thus,
for example, it provides the foundations for an equilibrium analysis of asset
markets in the presence of ambiguous volatility. The reader is referred to
\cite{ej} for further discussion of the framework and its economic
rationale.\footnote{For example, an occasional reaction is that ambiguity
about volatility is implausible because one can estimate the law of motion for
volatility (of asset prices, for example) extremely well. However, this
perspective presumes a stationary environment and relies on a tight connection
between the past and future that we relax. For similar reasons we reject the
suggestion that one can discriminate readily between nonequivalent laws of
motion and hence that there is no loss of empirical relevance in restricting
priors to be equivalent. See the companion paper for elaboration on these and
other matters of interpretation.}

The challenge in formulating the model is that it cannot be done within a
probability space framework.\footnote{Where the set of priors is finite (or
countable), a dominating measure is easily constructed. However, the set of
priors in our model is not countable, and a suitable probability space
framework does not exist outside of the extreme case where there is no
ambiguity about volatility.} Typically, the ambient framework is a probability
space $\left(  \Omega,P_{0}\right)  $, it is assumed that $B=\left(
B_{t}\right)  $ is a Brownian motion under $P_{0}$, and importantly, $P_{0}$
is used to define null events. Thus random variables and stochastic processes
are defined only up to the $P_{0}$-almost sure qualification and $P_{0}$ is an
essential part of the definition of all formal domains. However, ambiguity
about volatility implies that nullity (or possibility) cannot be defined by
any single probability measure. This is easily illustrated. Let $B$ be a
Brownian motion under $P_{0}$ and denote by $P^{\underline{\sigma}}$ and
$P^{\overline{\sigma}}$ the probability distributions over continuous paths
induced by $P_{0}$ and the two processes $($\underline{$\sigma$}$B_{t})$ and
$\left(  \overline{\sigma}B_{t}\right)  $, where, for simplicity,
\underline{$\sigma$} and $\overline{\sigma}$ are constants. Then
$P^{\underline{\sigma}}$ and $P^{\overline{\sigma}}$ are mutually singular
(and hence not equivalent) because
\begin{equation}
P^{\underline{\mathbf{\sigma}}}(\{\langle B\rangle_{T}=\underline{\sigma}%
^{2}T\})=1=P^{\overline{\mathbf{\sigma}}}(\{\langle B\rangle_{T}%
=\overline{\sigma}^{2}T\})\text{.}\label{Psingular}%
\end{equation}
To overcome the resulting difficulty, we define appropriate domains of
stochastic processes by using the entire set of priors to define the almost
sure qualification. For example, equality of two random variables will be
taken to mean almost sure equality for every prior in the decision maker's set
of priors. This so-called \emph{quasisure stochastic analysis} was developed
by Denis and Martini \cite{DM}. See also Soner et al. \cite{STZ} and Denis et
al. \cite{DHP} for elaboration on why a probability space framework is
inadequate and for a comparison of quasisure analysis with related
approaches.\footnote{Related developments are provided by Bion-Nadal et al.
\cite{bioan} \ and Soner et al. \cite{STZ-2010-2,STZ-2010-3}.} Prominent among
the latter is the seminal contribution of G-expectation due to Peng
\cite{P-2004,P-2006,P-2008,P-2010}, wherein a nonlinear expectations operator
is defined by a set of undominated measures. We combine elements of both
quasisure analysis and G-expectation. Conditioning, or updating, is obviously
a crucial ingredient in modeling dynamic preferences. In this respect we adapt
the approach in Soner et al. \cite{STZ-2010-4} and Nutz \cite{nutz} to
conditioning undominated sets of measures.\footnote{Peng \cite{P-2004,P-2005}
provides a related approach to conditioning.} However, these analyses do not
apply off-the-shelf because, for example, they permit ambiguity about
volatility but not about drift. In particular, accommodating both kinds of
ambiguity necessitates a novel construction of the set of priors.

Besides those already mentioned, there are only a few relevant papers in the
literature on continuous time utility theory. Denis and Kervarec
\cite{deniskervarec} formulate multiple priors utility functions and study
optimization in a continuous-time framework; they do not assume equivalence of
measures but they restrict attention to the case where only terminal
consumption matters and where all decisions are made at a single ex ante
stage. Bion-Nadal and Kervarec \cite{bioan} study risk measures (which can be
viewed as a form of utility function) in the absence of certainty about what
is possible.

Section \ref{section-utility} is the heart of the paper and presents the model
of utility, beginning with the construction of the set of priors and the
definition of (nonlinear) conditional expectation. Proofs are collected in appendices.

\section{Utility\label{section-utility}}

\subsection{Preliminaries\label{section-prelim}}

Time $t$ varies over the finite horizon $[0,T]$. Paths or trajectories of the
driving process are assumed to be continuous and thus are modeled by elements
of $C^{d}([0,T])$, the set of all $\mathbb{R}^{d}$-valued continuous functions
on $[0,T]$, endowed with the sup norm. The generic path is $\omega=(\omega
_{t})_{t\in\lbrack0,T]}$, where we write $\omega_{t}$ instead of
$\omega\left(  t\right)  $. All relevant paths begin at $0$ and thus we define
the canonical state space to be
\[
\Omega=\left\{  \omega=\left(  \omega_{t}\right)  \in C^{d}([0,T]):\omega
_{0}=0\right\}  \text{.}%
\]

The coordinate process $\left(  B_{t}\right)  $, where $B_{t}(\omega
)=\omega_{t}$,\ is denoted by $B$. Information is modeled by the filtration
$\mathcal{F}=\{\mathcal{F}_{t}\}$ generated by $B$. Let $P_{0}$ be the Wiener
measure on $\Omega$ so that $B$ is a Brownian motion under $P_{0}$. It is a
reference measure only in the mathematical sense of facilitating the
description of the individual's set of priors; but the latter need not contain
$P_{0}$.

Define also the set of paths over the time interval $[0,t]$: {\Large \ }%
\[
^{t}\Omega=\left\{  ^{t}\omega=\left(  ^{t}\omega_{s}\right)  \in
C^{d}([0,t]):^{t}\omega_{0}=0\right\}  \text{.}%
\]
Identify $^{t}\Omega$ with a subspace of $\Omega$ by identifying any
$^{t}\omega$ with the function on $[0,T]${\Large \ }that is constant at level
$^{t}\omega_{t}$ on $[t,T]$. Note that the filtration $\mathcal{F}_{t} $ is
the Borel $\sigma$-field on $^{t}\Omega$. (Below, for any topological space we
always adopt the Borel $\sigma$-field even where not mentioned explicitly.)

Consumption processes $c$ take values in $C$, a convex subset of
$\mathbb{R}^{\ell}$. The domain of consumption processes is denoted $D$.
Because we are interested in describing dynamic choice, we need to specify not
only ex ante utility over $D$, but a suitable process of (conditional) utility functions.

The key is construction of the set of priors. The primitive is the
individual's hypotheses about drift and volatility. These are used to specify
the set of priors. Conditioning is treated next. Finally, these components are
applied to define a recursive process of utility functions.

\subsection{Drift and Volatility Hypotheses\label{section-priors}}

Before moving to the general setup, we outline briefly a special case where
$d=1$ and there is ambiguity only about volatility. Accordingly, suppose that
(speaking informally) the individual is certain that the driving process
$B=(B_{t})$ is a martingale, but that its volatility is known only up to the
interval $\left[  \underline{\sigma},\overline{\sigma}\right]  $. In
particular, she does not take a stand on any particular parametric model of
volatility dynamics.

To be more precise about the meaning of volatility, recall that the quadratic
variation process of $(B_{t})$ is defined by%
\begin{equation}
\langle B\rangle_{t}(\omega)=\underset{\bigtriangleup t_{k}\rightarrow0}{\lim
}~\underset{t_{k}\leq t}{\Sigma}\mid B_{t_{k+1}}(\omega)-B_{t_{k}}(\omega
)\mid^{2}\label{Blim}%
\end{equation}
where $0=t_{1}<\ldots<t_{n}=t$ and $\bigtriangleup t_{k}=t_{k+1}-t_{k}$. (By
Follmer \cite{follmer} and Karandikar \cite{ka}, the above limit exists almost
surely for every measure that makes $B$ a martingale, thus giving a universal
stochastic process $\langle B\rangle$; because the individual is certain that
$B$ is a martingale, this limited universality is all we need.) Then the
volatility $\left(  \sigma_{t}\right)  $ of $B$ is defined by
\[
d\langle B\rangle_{t}=\sigma_{t}^{2}dt\text{.}%
\]
Therefore, the interval constraint on volatility can be written also in the
form
\begin{equation}
\underline{\sigma}^{2}t\leq\langle B\rangle_{t}\leq\overline{\sigma}%
^{2}t\text{.}\label{sigma-ineq}%
\end{equation}

The model that follows is much more general. Importantly, the interval
$\left[  \underline{\sigma},\overline{\sigma}\right]  $ can be time and state
varying, and the dependence on history of the interval at time $t$ is
unrestricted, thus permitting any model of how ambiguity varies with
observation (that is, learning) to be accommodated. In addition, the model
admits multidimensional driving processes ($d>1$) and also ambiguity about
drift, thus relaxing the assumption of certainty that $B$ is a martingale.

In general, the individual is not certain that the driving process has zero
drift and/or unit variance. Accordingly, she entertains a range of alternative
hypotheses $X^{\theta}=(X_{t}^{\theta})$ parametrized by $\theta=\left(
\theta_{t}\right)  $. Here $\theta_{t}=(\mu_{t},\sigma_{t}) $ is an
$\mathcal{F}$-progressively measurable process with values in $\mathbb{R}%
^{d}\times\mathbb{R}^{d\times d}$ that describes a conceivable process for
drift $\mu=\left(  \mu_{t}\right)  $ and for volatility $\sigma=\left(
\sigma_{t}\right)  $. The primitive is the process of correspondences
$(\Theta_{t})$, where, for each $t$,
\[
\Theta_{t}:\Omega\rightsquigarrow\mathbb{R}^{d}\times\mathbb{R}^{d\times
d}\text{.}%
\]
Roughly, $\Theta_{t}\left(  \omega\right)  $ gives the set of admissible drift
and volatility pairs at $t$ along the trajectory $\omega$. The idea is that
each $\theta$ parametrizes the driving process $X^{\theta}=(X_{t}^{\theta})$
given by the unique solution to the following stochastic differential equation
(SDE) under $P_{0}$:%
\begin{equation}
dX_{t}^{\theta}=\mu_{t}(X_{\cdot}^{\theta})dt+\sigma_{t}(X_{\cdot}^{\theta
})dB_{t}\text{,}\;\ \ X_{0}^{\theta}=0\text{,}\;t\in\lbrack0,T]\text{.}%
\label{Xtheta}%
\end{equation}
We assume that only $\theta$'s for which a unique strong solution exists are
adopted as hypotheses. Therefore, denote by $\Theta^{SDE}$ the set of all
processes $\theta$ that ensure a unique strong solution $X^{\theta}$ to the
SDE,\footnote{By uniqueness we mean that $P_{0}\left(  \left\{  \sup_{0\leq
t\leq T}\mid X_{t}^{\theta}-X_{t}^{\prime}\mid>0\right\}  \right)  =0$ for any
other strong solution $\left(  X_{t}^{\prime}\right)  $. A Lipschitz condition
and boundedness are sufficient for existence of a unique solution, but these
properties are not necessary and a Lipschitz condition is not imposed.} and
define the set $\Theta$ of admissible drift and volatility processes by%
\begin{equation}
\Theta=\left\{  \theta\in\Theta^{SDE}:\theta_{t}\left(  \omega\right)
\in\Theta_{t}\left(  \omega\right)  \text{ \ for all }\left(  t,\omega\right)
\in\lbrack0,T]\times\Omega\right\}  \text{.}\label{theta}%
\end{equation}

We impose the following technical regularity conditions on $(\Theta_{t})$:

\begin{enumerate}
\item[(i)] \textit{Measurability}: The correspondence $(t,\omega
)\longmapsto\Theta_{t}(\omega)$ on $[0,s]\times\Omega$ is $\mathcal{B}%
([0,s])\times\mathcal{F}_{s}$-measurable for every $0<s\leq T$.

\item[(ii)] \textit{Uniform Boundedness}: There is a compact subset
$\mathcal{K}$ in $\mathbb{R}^{d}\times\mathbb{R}^{d\times d}$ such that
$\Theta_{t}:\Omega\rightsquigarrow\mathcal{K}$ each $t$.

\item[(iii)] \textit{Compact-Convex}: Each $\Theta_{t}$ is compact-valued and convex-valued.

\item[(iv)] \textit{Uniform Nondegeneracy}: There exists $\hat{a}$, a $d\times
d$ real-valued positive definite matrix, such that for every $t$ and $\omega$,
if $\left(  \mu_{t},\sigma_{t}\right)  \in$ $\Theta_{t}(\omega) $, then
$\sigma_{t}\sigma_{t}^{\top}\geq\hat{a}$.

\item[(v)] \textit{Uniform Continuity}: The process $(\Theta_{t})$\ is
uniformly continuous in the sense defined in Appendix \ref{app-uniform}.

\item[(vi)] \textit{Uniform Interiority}: There exists{\Large \ }$\delta>0$
such that $ri^{\delta}\Theta_{t}(\omega)\not =\varnothing$ ~for all $t$ and
$\omega$, where $ri^{\delta}\Theta_{t}(\omega)$\ is the $\delta$-relative
interior of $\Theta_{t}(\omega)$. (For any $D\subseteq(\mathbb{R}^{d}%
\times\mathbb{R}^{d\times d})$\ and $\delta>0$, $ri^{\delta}D\equiv\{x\in
D:(x+B_{\delta}(x))\cap($aff $D)\subset D\}$, where aff $D$ is the affine hull
of $D$ and $B_{\delta}(x)$\ denotes the open ball of radius $\delta$.)

\item[(vii)] \textit{Uniform Affine Hull}: The affine hulls of $\Theta
_{t^{\prime}}(\omega^{\prime})$ and $\Theta_{t}(\omega)$ are the same for
every $(t^{\prime},\omega^{\prime})$ and $(t,\omega)$ in $[0,T]\times\Omega$.
\end{enumerate}

Conditions (i)-(iii) parallel assumptions made by Chen and Epstein \cite{CE}.
A form of Nondegeneracy is standard in financial economics. The remaining
conditions are adapted from Nutz \cite{nutz} and are imposed in order to
accommodate ambiguity in volatility. The major differences from Nutz'
assumptions are in (vi) and (vii). Translated into our setting, he assumes
that $int^{\delta}\Theta_{t}(\omega)\not =\varnothing$, where, for any $D$,
$int^{\delta}\Theta_{t}(\omega)=$ $\{x\in D:(x+B_{\delta}(x))\subset D\}$. By
weakening his requirement to deal with relative interiors, we are able to
broaden the scope of the model in important ways (see the first three examples
below). Because each $\Theta_{t}(\omega)$ is convex, if it also has nonempty
interior then its affine hull is all of $\mathbb{R}^{d}\times\mathbb{R}%
^{d\times d}$. Then $ri^{\delta}\Theta_{t}(\omega)=int^{\delta}\Theta
_{t}(\omega)\not =\varnothing$ and also (vii) is implied. In this sense,
(vi)-(vii) are jointly weaker than assuming $int^{\delta}\Theta_{t}%
(\omega)\not =\varnothing$.

We illustrate the scope of the model through some examples.

\begin{example}
[Ambiguous drift]\label{example-drift}If $\Theta_{t}\left(  \omega\right)
\subset\mathbb{R}^{d}\times\{\sigma_{t}\left(  \omega\right)  \}$ for every
$t$ and $\omega$, for some volatility process $\sigma$, then there is
ambiguity only about drift. If $d=1$, it is modeled by the random and time
varying interval $[$\underline{$\mu$}$_{t},\overline{\mu}_{t}]$. The
regularity conditions above for $(\Theta_{t})$ are satisfied if: $\sigma
_{t}^{2}\geq a>0$ and $\overline{\mu}_{t}-$\underline{$\mu$}$_{t}>0$
everywhere, and if $\overline{\mu}_{t}$ and \underline{$\mu$}$_{t}$ are
continuous in $\omega$ uniformly in $t$. This special case corresponds to the
Chen and Epstein \cite{CE} model.
\end{example}

\begin{example}
[Ambiguous volatility]\label{example-vol}If $\Theta_{t}\left(  \omega\right)
\subset\{\mu_{t}\left(  \omega\right)  \}\times\mathbb{R}^{d\times d}$ for
every $t$ and $\omega$, for some drift process $\mu$, then there is ambiguity
only about volatility. If $d=1$, it is modeled by the random and time varying
interval $[$\underline{$\sigma$}$_{t},\overline{\sigma}_{t}]$. The regularity
conditions for $(\Theta_{t})$ are satisfied if: $\overline{\sigma}_{t}%
>$\underline{$\sigma$}$_{t}\geq a>0$ everywhere, and if $\overline{\sigma}%
_{t}$ and \underline{$\sigma$}$_{t}$ are continuous in $\omega$ uniformly in
$t$.

A generalization is important. Allow $d\geq1$ and let $\mu_{t}=0$. Then
(speaking informally) there is certainty that $B$ is a martingale in spite of
uncertainty about the true probability law. Volatility $\left(  \sigma
_{t}\right)  $ is a process of $d\times d$ matrices. Let the admissible
volatility processes $\left(  \sigma_{t}\right)  $ be those satisfying
$\sigma_{t}\in\Gamma$, where $\Gamma$ is any compact convex subset of
$\mathbb{R}^{d\times d}$ such that, for all $\sigma$ in $\Gamma$,
$\sigma\sigma^{\top}\geq\hat{a}$ for some positive definite matrix
$\widehat{a}$.\footnote{Given symmetric matrices $A^{\prime}$ and $A$,
$A^{\prime}\geq A$ if $A^{\prime}-A$ is positive semidefinite.} This
specification is essentially equivalent to Peng's \cite{P-2006} notion of
$G$\emph{-Brownian motion}.\footnote{Peng provides generalizations of
It\^{o}'s Lemma and It\^{o} integration appropriate for $G$-Brownian motion
that we exploit in our companion paper in deriving asset pricing
implications.}
\end{example}

\begin{example}
[Robust stochastic volatility]\label{example-robust}This is a special case of
the preceding example but we describe it separately in order to highlight the
connection of our model to the stochastic volatility literature. By a
stochastic volatility model we mean the hypothesis that the driving process
has zero drift and that its volatility is stochastic and is described by a
single process $\left(  \sigma_{t}\right)  $ satisfying regularity conditions
of the sort given above. The specification of a single process for volatility
indicates the individual's complete confidence in the implied dynamics.
Suppose, however, that $\left(  \sigma_{t}^{1}\right)  $ and $\left(
\sigma_{t}^{2}\right)  $ describe two alternative stochastic volatility models
that are put forth by expert econometricians;\footnote{There is an obvious
extension to any finite number of models.} for instance, they might conform to
the Hull and White \cite{hw} and Heston \cite{heston} parametric forms
respectively. The models have comparable empirical credentials and are not
easily distinguished empirically, but their implications for optimal choice
(or for the pricing of derivative securities, which is a context in which
stochastic volatility models are used heavily) differ significantly. Faced
with these two models, the individual might place probability $\frac{1}{2}$ on
each being the true model. But why should she be certain that either one is
true? Both $\left(  \sigma_{t}^{1}\right)  $ and $\left(  \sigma_{t}%
^{2}\right)  $ may fit data well to some approximation, but other
approximating models may do as well. An intermediate model such as $\left(
\frac{1}{2}\sigma_{t}^{1}+\frac{1}{2}\sigma_{t}^{2}\right)  $ is one
alternative, but there are many others that \textquotedblleft lie
between\textquotedblright\ $\left(  \sigma_{t}^{1}\right)  $ and $\left(
\sigma_{t}^{2}\right)  $ and that plausibly should be taken into account.
Accordingly, (assuming $d=1$), let%
\[
\underline{\sigma}_{t}\left(  \omega\right)  =\min\{\sigma_{t}^{1}\left(
\omega\right)  ,\sigma_{t}^{2}\left(  \omega\right)  \}\text{ and }%
\overline{\sigma}_{t}\left(  \omega\right)  =\max\{\sigma_{t}^{1}\left(
\omega\right)  ,\sigma_{t}^{2}\left(  \omega\right)  \}\text{,}%
\]
and admit all volatility processes with values lying in the interval
$[\underline{\sigma}_{t}\left(  \omega\right)  ,\overline{\sigma}_{t}\left(
\omega\right)  ]$ \ for every $\omega$.
\end{example}

\begin{example}
[Joint ambiguity]\label{example-joint}The model is flexible in the way it
relates ambiguity about drift and ambiguity about volatility. For example, a
form of independence is modeled if (taking $d=1$)
\begin{equation}
\Theta_{t}(\omega)=[\underline{\mu}_{t}\left(  \omega\right)  ,\overline{\mu
}_{t}\left(  \omega\right)  ]\times\lbrack\underline{\sigma}_{t}\left(
\omega\right)  ,\overline{\sigma}_{t}\left(  \omega\right)  ]\text{,}%
\label{indep}%
\end{equation}
the Cartesian product of the intervals described in the preceding two examples.

An alternative hypothesis is that drift and volatility are thought to move
together. This is captured by specifying, for example,
\begin{equation}
\Theta_{t}(\omega)=\{(\mu,\sigma)\in\mathbb{R}^{2}:\mu=\mu_{\min}%
+z,~\sigma^{2}=\sigma_{\min}^{2}+2z/\gamma,~0\leq z\leq\overline{z}_{t}\left(
\omega\right)  \} \text{,}\label{joint}%
\end{equation}
where $\mu_{\min}$, $\sigma_{\min}^{2}$ and $\gamma>0$ are fixed and known
parameters. The regularity conditions for $(\Theta_{t})$ are satisfied if
$\overline{z}_{t}$ is positive everywhere and continuous in $\omega$ uniformly
in $t$. This specification is adapted from Epstein and Schneider \cite{esAR}.
\end{example}

\begin{example}
[Markovian ambiguity]\label{example-markov}Assume that $(\Theta_{t})$
satisfies:%
\[
\omega_{t}^{\prime}=\omega_{t}~\Longrightarrow~\Theta_{t}\left(
\omega^{\prime}\right)  =\Theta_{t}\left(  \omega\right)  \text{.}%
\]
Then ambiguity depends only on the current state and not on history. Note,
however, that according to (\ref{theta}), the drift and volatility processes
deemed possible are not necessarily Markovian - $\theta_{t}$ can depend on the
complete history at any time. Thus the individual is \emph{not} certain that
the driving process is Markovian, but the \emph{set} of processes that she
considers possible at any given time is independent of history beyond the
prevailing state.
\end{example}

\subsection{Priors, expectation and conditional expectation}

We proceed to translate the set $\Theta$ of hypotheses about drift and
volatility into a set of priors. Each $\theta$ induces (via $P_{0}$) a
probability measure $P^{\theta}$ on $(\Omega,\mathcal{F}_{T})$ given by
\[
P^{\theta}(A)=P_{0}(\{\omega:X^{\theta}(\omega)\in A)\text{, }A\in
\mathcal{F}_{T}\text{.}%
\]
Therefore, we arrive at the set of priors $\mathcal{P}^{\Theta}$ given by
\begin{equation}
\mathcal{P}^{\Theta}\mathcal{=\{}P^{\theta}:\theta\in\Theta\}.\label{Ptheta}%
\end{equation}
Fix $\Theta$ and denote the set of priors $\mathcal{P}^{\Theta}$ simply by
$\mathcal{P}$. This is the set of priors used, as in the Gilboa-Schmeidler
model, to define utility and to describe choice between consumption
processes.\footnote{The set $\mathcal{P}$ is relatively compact in the
topology induced by bounded continuous functions (this is a direct consequence
of Gihman and Skorohod \cite[Theorem 3.10]{GSb}).}

\begin{remark}
If all alternative hypotheses display unit variance, then ambiguity is limited
to the drift as in the Chen-Epstein model, and one can show that, as in
\cite{CE}, measures in $\mathcal{P}$ are pairwise equivalent. At the other
extreme, if they all display zero drift, then ambiguity is limited to
volatility and many measures in $\mathcal{P}$ are mutually singular.
Nonequivalence of priors prevails in the general model when both drift and
volatility are ambiguous.
\end{remark}

Given $\mathcal{P}$, we define (nonlinear) expectation as follows. For a
random variable $\xi$\ on $(\Omega,\mathcal{F}_{T})$, if $\sup_{P\in
\mathcal{P}}E_{P}\xi<\infty$ define
\begin{equation}
\hat{E}\xi=\sup_{P\in\mathcal{P}}E_{P}\xi.\label{Ehat}%
\end{equation}
Because we will assume that the individual is concerned with worst-case
scenarios, below we use the fact that%
\[
\inf_{P\in\mathcal{P}}E_{P}\xi=-\hat{E}[-\xi]\text{.}%
\]

The crucial remaining ingredient of the model, and the focus of most of the
work in the appendices, is conditioning. A naive approach to defining
conditional expectation would be to use the standard conditional expectation
$E_{P}[\xi\mid\mathcal{F}_{t}]$ for each $P$ in $\mathcal{P}$ and then to take
the (essential) supremum over $\mathcal{P}$. Such an approach immediately
encounters a roadblock due to the nonequivalence of priors. The conditional
expectation $E_{P}[\xi\mid\mathcal{F}_{t}]$ is well defined only $P$-almost
surely, while to be a meaningful object for analysis, a random variable must
be well defined from the perspective of every measure in $\mathcal{P}$. In the
following, we say that a property holds \emph{quasisurely} ($q.s.$\ for short)
if it holds $P$-$a.s.$\ for every $P\in\mathcal{P}$.\footnote{Throughout, when
$Z$ is a random variable, $Z\geq0$ quasisurely means that the inequality is
valid $P$-$a.s.$ for every $P$ in $P$. If $Z=\left(  Z_{t}\right)  $ is a
process, \noindent by the statement \textquotedblleft$Z_{t}\geq0$ for every
$t$ quasisurely (q.s.)\textquotedblright\ we mean that for every $t$ there
exists $G_{t}\subset\Omega$ such that $Z_{t}\left(  \omega\right)  \geq0$ for
all $\omega\in G_{t}$ and $P\left(  G_{t}\right)  =1$ for all $P$ in $P$. If
$Z_{t}=0$ for every $t$\ quasisurely, then $Z=0$ in $M^{2}(0,T)$ (because
$\hat{E}[%
{\displaystyle\int\nolimits_{0}^{T}}
\mid Z_{t}\mid^{2}dt]\leq%
{\displaystyle\int\nolimits_{0}^{T}}
\hat{E}[\mid Z_{t}\mid^{2}]dt$), but the converse is not valid in
general.}\ In other words, and speaking very informally, conditional beliefs
must be defined at every node deemed possible by some measure in $\mathcal{P}%
$. The economic rationale is that even if $P\left(  A\right)  =0$, for some
$A\in\mathcal{F}_{t}$ and $t>0$, if also $Q\left(  A\right)  >0$ for some
other prior in $\mathcal{P}$, then ex ante the individual does not totally
dismiss the possibility of $A$ occurring when she formulates consumption
plans: if she is guided by the worst case scenario, then $\min_{P^{\prime}%
\in\mathcal{P}}P^{\prime}\left(  \Omega\backslash A\right)  <1$ implies that
she would reject a bet against $A$ that promised a sufficiently poor prize
(low consumption stream) if $A$ occurs. Therefore, a model of dynamic choice
by a sophisticated and forward-looking individual should specify her
consumption plan contingent on arriving at $(A,t)$.

This difficulty can be overcome because for every admissible hypothesis
$\theta$, $\theta_{t}\left(  \omega\right)  $ is defined for \emph{every}
$\left(  t,\omega\right)  $, that is, the primitives specify a hypothesized
instantaneous drift-volatility pair everywhere in the tree. This feature of
the model resembles the approach adopted in the theory of extensive form
games, namely the use of conditional probability systems, whereby conditional
beliefs at \emph{every} node are specified as primitives, obviating the need
to update. It resembles also the approach in the discrete time model in
Epstein and Schneider \cite{es2003}, where roughly, conditional beliefs about
the next instant for every time and history are adopted as primitives and are
pasted together by backward induction to deliver the ex ante set of priors.

To proceed, recall the construction of the set of priors through
(\ref{Xtheta}) and the set $\Theta$ of admissible drift and volatility
processes. If $\theta=\left(  \theta_{s}\right)  $ is a conceivable scenario
ex ante, then $(\theta_{s}\left(  ^{t}\omega,\cdot\right)  )_{t\leq s\leq T} $
$\ $is seen by the individual ex ante as a conceivable continuation from time
$t$ along the history $\omega$. We assume that then it is also a conceivable
scenario ex post conditionally on $\left(  t,\omega\right)  $, thus ruling out
surprises or unanticipated changes in outlook. Accordingly, $X^{\theta
,t,\omega}=(X_{s}^{\theta,t,\omega})_{t\leq s\leq T}$ is a conceivable
conditional scenario for the driving process if it solves the following SDE
under $P_{0}$:%
\begin{equation}
\left\{
\begin{array}
[c]{rl}%
dX_{s}^{\theta,t,\omega} & =\mu_{s}(X_{\cdot}^{\theta,t,\omega})ds+\sigma
_{s}(X_{\cdot}^{\theta,t,\omega})dB_{s},\;t\leq s\leq T\\
X_{s}^{\theta,t,\omega} & =\omega_{s},\;0\leq s\leq t\text{.}%
\end{array}
\right. \label{Xrcp}%
\end{equation}
The solution $X^{\theta,t,\omega}$ induces a probability measure
$P_{t}^{\theta,\omega}\in\Delta\left(  \Omega\right)  $, denoted simply by
$P_{t}^{\omega}$ with $\theta$ suppressed when it is understood that
$P=P^{\theta}$. For each $P$ in $\mathcal{P}$, the measure $P_{t}^{\omega}%
\in\Delta\left(  \Omega\right)  $ is defined for every $t$ and $\omega$, and,
importantly, it is a version of the regular $\mathcal{F}_{t}$-conditional
probability of $P$ (see Lemma \ref{lemma-rcp}).

The set of all such conditionals obtained as $\theta$ varies over $\Theta$ is
denoted $\mathcal{P}_{t}^{\omega}$, that is,
\begin{equation}
\mathcal{P}_{t}^{\omega}=\left\{  P_{t}^{\omega}:P\in\mathcal{P}\right\}
\text{.}\label{regcond}%
\end{equation}
We take $\mathcal{P}_{t}^{\omega}$ to be the individual's set of priors
conditional on $\left(  t,\omega\right)  $.\footnote{The evident parallel with
the earlier construction of the ex ante set $\mathcal{P}$ can be expressed
more formally because the construction via (\ref{Xrcp}) can be expressed in
terms of a process of correspondences $\left(  \Theta_{s}^{t,\omega}\right)
_{t\leq s\leq T}$, $\Theta_{s}^{t,\omega}:C^{d}([t,T])\rightsquigarrow
\mathbb{R}^{d}\times\mathbb{R}^{d\times d}$, satisfying counterparts of the
regularity conditions (i)-(vii) on the time interval $[t,T]$.}

The sets of conditionals in (\ref{regcond}) lead to the following (nonlinear)
conditional expectation on $UC_{b}\left(  \Omega\right)  $, the set of all
bounded and uniformly continuous functions on $\Omega$:
\begin{equation}
\hat{E}[\xi\mid\mathcal{F}_{t}]\left(  \omega\right)  =\sup_{P\in
\mathcal{P}_{t}^{\omega}}E_{P}\xi\text{, \ for every }\xi\in UC_{b}\left(
\Omega\right)  \text{ and }(t,\omega)\in\lbrack0,T]\times\Omega
.\label{Econditional2}%
\end{equation}
Conditional expectation is defined thereby on $UC_{b}\left(  \Omega\right)  $,
but this domain is not large enough for our purposes.\footnote{The definition
is restricted to $UC_{b}\left(  \Omega\right)  $ in order to ensure the
measurability of $\hat{E}[\xi\mid F_{t}]\left(  \cdot\right)  $ and proof of a
suitable form of the law of iterated expectations.\ Indeed, Appendix
\ref{app-conditioning}, specifically (\ref{Econditional}), shows that
conditional expectation has a different representation when random variables
outside $UC_{b}\left(  \Omega\right)  $ are considered.} For example, the
conditional expectation of $\xi$ need not be (uniformly) continuous in
$\omega$ even if $\xi$ is bounded and uniformly continuous, which is an
obstacle to dealing with stochastic processes and recursive modeling.
(Similarly, if one were to use the space $C_{b}\left(  \Omega\right)  $ of
bounded continuous functions.)

Thus we consider the larger domain $\widehat{L^{2}}(\Omega)$, the completion
of $UC_{b}\left(  \Omega\right)  $\ under the norm $\parallel\xi
\parallel\equiv(\hat{E}[\mid\xi\mid^{2}])^{\frac{1}{2}}$.\footnote{It
coincides with the completion of $C_{b}\left(  \Omega\right)  $; see
\cite{DHP}.} Denis et al. \cite{DHP} show that a random variable $\xi$ defined
on $\Omega$ lies in $\widehat{L^{2}}(\Omega)$ if and only if: (i) $\xi$ is
\emph{quasicontinuous} - for every $\epsilon>0$ there exists an open set
$G\subset\Omega$ with $P\left(  G\right)  <\epsilon$ for every $P$ in
$\mathcal{P}$ such that $\xi$ is continuous on $\Omega\backslash G$; and (ii)
$\xi$ is uniformly integrable in the sense that $\lim_{n\rightarrow\infty}%
\sup_{P\in\mathcal{P}}E^{P}\left(  \mid\xi\mid^{2}\mathbf{1}_{\{\mid\xi
\mid>n\}}\right)  =0$. This characterization is inspired by Lusin's Theorem
for the classical case which implies that when $\mathcal{P}=\{P\}$, then
$\widehat{L^{2}}(\Omega)$ reduces to the familiar space of $P$-squared
integrable random variables. For general $\mathcal{P}$, $\widehat{L^{2}%
}(\Omega)$ is a proper subset of the set of measurable random variables $\xi$
for which $\sup_{P\in\mathcal{P}}E^{P}\left(  \mid\xi\mid^{2}\right)  <\infty
$.\footnote{For example, if $\mathcal{P}$ is the set of all Dirac measures
with support in $\Omega$, then $\widehat{L^{2}}(\Omega)=C_{b}\left(
\Omega\right)  $.} However, it is large in the sense of containing many
discontinuous random variables; for example, $\widehat{L^{2}}(\Omega)$
contains every bounded and lower semicontinuous function on $\Omega$ (see the
proof of Lemma \ref{lemma-extension}).

Another aspect of $\widehat{L^{2}}(\Omega)$ warrants emphasis. Two random
variables $\xi^{\prime}$ and $\xi$ are identified in $\widehat{L^{2}}(\Omega)$
if and only if $\parallel\xi^{\prime}-\xi\parallel=0$, which means that
$\xi^{\prime}=\xi$ almost surely with respect to $P$ for \emph{every} $P$ in
$\mathcal{P}$. In that case, say that the equality obtains quasisurely and
write $\xi^{\prime}=\xi$ $\ q.s.$ Thus $\xi^{\prime}$ and $\xi$ are
distinguished whenever they differ with positive probability for \emph{some}
measure in $\mathcal{P}$. Accordingly, the space $\widehat{L^{2}}(\Omega)$
provides a more detailed picture of random variables than does any single
measure in $\mathcal{P}$.\

The next theorem (proven in Appendix \ref{app-conditioning}) shows that
conditional expectation admits a suitably unique and well behaved extension
from $UC_{b}\left(  \Omega\right)  $ to all of $\widehat{L^{2}}(\Omega)$.
Accordingly, the sets $\mathcal{P}_{t}^{\omega}$ determine the conditional
expectation for all random variables considered in the sequel.

\begin{theorem}
[Conditioning]\label{thm-conditioning}The mapping $\hat{E}[\cdot
\mid\mathcal{F}_{t}]$ on $UC_{b}(\Omega)$ defined in (\ref{Econditional2}) can
be extended uniquely to a $1$-Lipschitz continuous mapping $\hat{E}[\cdot
\mid\mathcal{F}_{t}]:\widehat{L^{2}}(\Omega)\rightarrow\widehat{L^{2}}%
(^{t}\Omega)$, where $1$-Lipschitz continuity means that%
\[
\parallel\hat{E}[\xi^{\prime}\mid\mathcal{F}_{t}]-\hat{E}[\xi\mid
\mathcal{F}_{t}]\parallel_{\widehat{L^{2}}}\leq\parallel\xi^{\prime}%
-\xi\parallel_{\widehat{L^{2}}}\text{ \ for all }\xi^{\prime},\xi
\in\widehat{L^{2}}(\Omega).
\]
Moreover, the extension satisfies, for all $\xi$ and $\eta$ in $\widehat{L^{2}%
}(\Omega)$\ and for all $t\in\lbrack0,T]$,
\begin{equation}
\hat{E}[\hat{E}[\xi\mid\mathcal{F}_{t}]\mid\mathcal{F}_{s}]=\hat{E}[\xi
\mid\mathcal{F}_{s}],\;\text{\ for }0\leq s\leq t\leq T\text{,}\label{LIE}%
\end{equation}
and:

(i) If $\xi\geq\eta$, then $\hat{E}[\xi\mid\mathcal{F}_{t}]\geq\hat{E}%
[\eta\mid\mathcal{F}_{t}]$.

(ii) If $\xi$\ is $\mathcal{F}_{t}$-measurable, then $\hat{E}[\xi
\mid\mathcal{F}_{t}]=\xi$.

(iii) $\hat{E}[\xi\mid\mathcal{F}_{t}]+\hat{E}[\eta\mid\mathcal{F}_{t}%
]\geq\hat{E}[\xi+\eta\mid\mathcal{F}_{t}]$ with equality if $\eta$ is
$\mathcal{F}_{t}$-measurable.

(iv) $\hat{E}[\eta\xi\mid\mathcal{F}_{t}]=\eta^{+}\hat{E}[\xi\mid
\mathcal{F}_{t}]+\eta^{-}\hat{E}[-\xi\mid\mathcal{F}_{t}]$, if $\eta$ is
$\mathcal{F}_{t}$-measurable.
\end{theorem}

The Lipschitz property is familiar from the classical case of a single prior,
where it is implied by Jensen's inequality (see the proof of Lemma
\ref{lemma-extension}). The law of iterated expectations (\ref{LIE}) is
intimately tied to dynamic consistency of the preferences discussed below. The
nonlinearity expressed in (iii) reflects the nonsingleton nature of the set of
priors. Other properties have clear interpretations.

\subsection{The definition of utility}

Because we turn to consideration of processes, we define $M^{2,0}(0,T)$, the
class of processes $\eta$ of the form%
\[
\eta_{t}(\omega)=%
{\displaystyle\sum\limits_{i=0}^{N-1}}
\xi_{i}(\omega)1_{[t_{i},t_{i+1})}(t),
\]
where $\xi_{i}\in\widehat{L^{2}}(^{t_{i}}\Omega)$, $0\leq i\leq N-1$, and
$0=t_{0}<\cdots<t_{N}=T$.\footnote{The space $^{t_{i}}\Omega$ was defined in
Section \ref{section-prelim}.} Roughly, each such $\eta$ is a step function in
random variables from the spaces $\widehat{L^{2}}(^{t_{i}}\Omega)$. For the
usual technical reasons, we consider also suitable limits of such processes.
Thus define $M^{2}(0,T)$ to be the completion of $M^{2,0}(0,T)$ under the norm%
\[
\parallel\eta\parallel_{M^{2}(0,T)}\equiv(\hat{E}[%
{\displaystyle\int\nolimits_{0}^{T}}
\mid\eta_{t}\mid^{2}dt])^{\frac{1}{2}}.
\]

Consumption at every time takes values in $C$, a convex subset of
$\mathbb{R}_{+}^{d}$. Consumption processes $c=(c_{t})$ lie in $D$, a subset
of $M^{2}(0,T)$.

For each $c$ in $D$, we define a utility process $\left(  V_{t}\left(
c\right)  \right)  $, where $V_{t}\left(  c\right)  $ is the utility of the
continuation $\left(  c_{s}\right)  _{0\leq s\leq t}$ and $V_{0}\left(
c\right)  $ is the utility of the entire process $c$. We often suppress the
dependence on $c$ and write simply $\left(  V_{t}\right)  $. We define utility
following Duffie and Epstein \cite{DE}. This is done in order that our model
retain the flexibility to partially separate intertemporal substitution from
other aspects of preference (here uncertainty, rather than risk by which we
mean `probabilistic uncertainty').

Let $\Theta$ and $\mathcal{P}=\mathcal{P}^{\Theta}$ be as above. The other
primitive component is the \emph{aggregator} $f:C\times\mathbb{R}%
^{1}\rightarrow\mathbb{R}^{1}$. It is assumed to satisfy:

(i) $f$ is Borel measurable.

(ii) Uniform Lipschitz for aggregator: There exists a positive constant $K$
such that
\[
\mid f(c,v^{\prime})-f(c,v)\mid\leq K\mid v^{\prime}-v\mid\text{,}\;\text{for
all }(c,v^{\prime},v)\in C\times\mathbb{R}^{2}.
\]

(iii) $(f(c_{t},v))_{0\leq t\leq T}\in M^{2}(0,T)$ for each $v\in R$ and $c\in
D$.

\bigskip

We define $V_{t}$ by%

\begin{equation}%
\begin{array}
[c]{cl}%
V_{t} & =-\hat{E}[-\int_{t}^{T}f(c_{s},V_{s})ds\mid\mathcal{F}_{t}]
\end{array}
\text{.}\label{Vt}%
\end{equation}
This definition of utility generalizes both Duffie and Epstein \cite{DE},
where there is no ambiguity, and Chen and Epstein \cite{CE}, where ambiguity
is confined to drift. Formally, they use different sets of priors: a singleton
set in the former paper and a suitable set of equivalent priors in the latter paper.

Our main result follows (see Appendix \ref{app-utility} for a proof).

\begin{theorem}
[Utility]\label{thm-utility}Let $(\Theta_{t})$ and $f$ satisfy the above
assumptions. Fix $c\in D$. Then:

(a) There exists a unique process $\left(  V_{t}\right)  $ in $M^{2}(0,T)$
solving (\ref{Vt}).

(b) The process $\left(  V_{t}\right)  $ is the unique solution in
$M^{2}(0,T)$ to $V_{T}=0$ and%
\begin{equation}
V_{t}=-\, \widehat{E}\left[  -\int_{t}^{\tau}\,f(c_{s},V_{s})\,ds\,-\,V_{\tau
}\, \mid\, \mathcal{F}_{t}\right]  \text{, \ }\, \,0\leq t<\tau\leq
T.\label{Vrecursive}%
\end{equation}

\end{theorem}

Part (a) proves that utility is well defined by (\ref{Vt}). Recursivity is
established in (b).

The most commonly used aggregator has the form%
\begin{equation}
f\left(  c_{t},v\right)  =u\left(  c_{t}\right)  -\beta v\text{, ~\ }\beta
\geq0\text{,}\label{fstandard}%
\end{equation}
in which case utility admits the closed-form expression%
\begin{equation}
V_{t}=-\hat{E}[-\int_{t}^{T}u(c_{s})e^{-\beta s}ds\mid\mathcal{F}%
_{t}].\label{Vstandard}%
\end{equation}

More generally, closed form expressions are rare. The following example
illustrates the effect of volatility ambiguity.

\begin{example}
[Closed form]\label{example-closedform}Consider the consumption process $c$
satisfying (under $P_{0}$)%
\begin{equation}
d\log c_{t}=s^{\top}\sigma_{t}dB_{t}\text{, ~}c_{0}>0\text{ given,}\label{e}%
\end{equation}
where $s$ is constant and the volatility matrix $\sigma_{t}$ is restricted
only to lie in the compact and convex set $\Gamma$, as in Example
\ref{example-vol}, corresponding to Peng's \cite{P-2006} notion of
$G$-Brownian motion. Utility is defined by the standard aggregator,%
\[
V_{t}\left(  c\right)  =-\hat{E}[-\int_{t}^{T}u(c_{s})e^{-\beta(s-t)}%
ds\mid\mathcal{F}_{t}]\text{.}%
\]
where the felicity function $u$ is given by
\[
u\left(  c_{t}\right)  =(c_{t})^{\alpha}/\alpha\text{, ~}0\not =%
\alpha<1\text{.}%
\]
Then the conditional utilities $V_{t}\left(  c\right)  $ can be expressed in
closed form. To do so, define \underline{$\sigma$}\ and $\overline{\sigma}$ as
the respective solutions to%
\begin{equation}
\min_{\sigma\in\Gamma}tr\left(  \sigma\sigma^{\top}ss^{\top}\right)  \text{
and}\ \max_{\sigma\in\Gamma}tr\left(  \sigma\sigma^{\top}ss^{\top}\right)
.\label{sigmabar}%
\end{equation}
(If $d=1$, then $\Gamma$ is a compact interval and \underline{$\sigma$} and
$\overline{\sigma}$ are its left and right endpoints.)

Let $P^{\ast}$\ be the measure on $\Omega$\ induced by $P_{0}$ and $X^{\ast}$,
where%
\[
X_{t}^{\ast}=\overline{\sigma}^{\top}B_{t}\text{, for all }t\text{ and }%
\omega\text{;}%
\]
define $P^{\ast\ast}$\ similarly using \underline{$\sigma$} and $X^{\ast\ast}%
$. They are worst-case (or minimizing) measures in that\footnote{That the
minimizing measure corresponds to constant volatility is a feature of this
example. More generally, the minimizing measure in $\mathcal{P}$ defines a
specific stochastic volatility model.}%
\begin{equation}
V_{0}\left(  c\right)  =\left\{
\begin{array}
[c]{cc}%
E^{P^{\ast}}\left[  \int_{0}^{T}\alpha^{-1}(c_{\tau})^{\alpha}e^{-\beta\tau
}d\tau\right]  & \text{if }\alpha<0\\
E^{P^{\ast\ast}}\left[  \int_{0}^{T}\alpha^{-1}(c_{\tau})^{\alpha}%
e^{-\beta\tau}d\tau\right]  & \text{if }\alpha>0
\end{array}
\right. \label{V0c}%
\end{equation}
This follows from Levy et al. \cite{levy} and Peng \cite{P-2010}, because
$u\left(  c_{t}\right)  =\alpha^{-1}\exp\left(  \alpha\log c_{t}\right)  $ and
$x\longmapsto e^{\alpha x}/\alpha$ is concave if $\alpha<0$ and convex if
$\alpha>0$. From (\ref{e}), almost surely with respect to $P^{\left(
\sigma_{\tau}\right)  }$,
\[
\alpha^{-1}c_{t}^{\alpha}=\alpha^{-1}c_{0}^{\alpha}\exp\left\{  \alpha\int%
_{0}^{t}s^{\top}\sigma_{\tau}dB_{\tau}\right\}  \text{.}%
\]
It follows that utility can be computed as if the volatility $\sigma_{t}$ were
constant and equal to $\overline{\sigma}$ (if $\alpha<0$) or $\ $%
\underline{$\sigma$}\ (if $\alpha>0$).

For $t>0$, employ the regular conditionals of $P^{\ast}$ and $P^{\ast\ast} $,
which have a simple form. For example, following (\ref{Xrcp}), for every
$\left(  t,\omega\right)  $, $(P^{\ast})_{t}^{\omega}$ is the measure on
$\Omega$ induced by the SDE
\[
\left\{
\begin{array}
[c]{rl}%
dX_{\tau} & =\overline{\sigma}dB_{\tau},\;t\leq\tau\leq T\\
X_{\tau} & =\omega_{\tau},\;0\leq\tau\leq t
\end{array}
\right.
\]
Thus under $(P^{\ast})_{t}^{\omega}$, $B_{\tau}-B_{t}$ is $N\left(
0,\overline{\sigma}\overline{\sigma}^{\top}\left(  \tau-t\right)  \right)  $
for $t\leq\tau\leq T$. Further, $(P^{\ast})_{t}^{\omega}$ is the worst case
measure in $\mathcal{P}_{t}^{\omega}$ if $\alpha<0$; similarly, $(P^{\ast\ast
})_{t}^{\omega}$ is the worst case measure in $\mathcal{P}_{t}^{\omega}$ if
$\alpha>0$. Repeat the argument used above for $t=0$ to obtain%
\[
V_{t}\left(  c\right)  =\left\{
\begin{array}
[c]{cc}%
E^{(P^{\ast})_{t}^{\omega}}\left[  \int_{t}^{T}\alpha^{-1}(c_{\tau})^{\alpha
}e^{-\beta(\tau-t)}d\tau\right]  & \text{if }\alpha<0\\
E^{(P^{\ast\ast})_{t}^{\omega}}\left[  \int_{t}^{T}\alpha^{-1}(c_{\tau
})^{\alpha}e^{-\beta(\tau-t)}d\tau\right]  & \text{if }\alpha>0
\end{array}
\right.
\]
and, by computing the expectations, that%
\[
V_{t}\left(  c\right)  =\left\{
\begin{array}
[c]{cc}%
\alpha^{-1}c_{t}^{\alpha}\bar{\gamma}^{-1}(1-e^{-\bar{\gamma}(T-t)}) &
\text{if }\alpha<0\\
\alpha^{-1}c_{t}^{\alpha}\underline{\gamma}^{-1}(1-e^{-\underline{\gamma
}(T-t)}) & \text{if }\alpha>0
\end{array}
\right.
\]
where
\[
\bar{\gamma}\triangleq\beta-\frac{1}{2}\alpha^{2}s^{\top}\overline{\sigma
}\overline{\sigma}^{\top}s\text{, }\underline{\gamma}\triangleq\beta-\frac
{1}{2}\alpha^{2}s^{\top}\underline{\sigma}\underline{\sigma}^{\top}s\text{.}%
\]

\end{example}

Utility has a range of natural properties. Most noteworthy is that the process
$\left(  V_{t}\right)  $ satisfies the recursive relation (\ref{Vrecursive}).
However, though such recursivity is typically thought to imply dynamic
consistency, the nonequivalence of priors complicates matters. The noted
recursivity implies the following weak form of dynamic consistency: For any
$0<t<T$, and any two consumption processes $c^{\prime}$ and $c$ that coincide
on $[0,t]$,
\[
\lbrack V_{t}\left(  c^{\prime}\right)  \geq V_{t}\left(  c\right)  \text{
\ }q.s.]\Longrightarrow V_{0}\left(  c^{\prime}\right)  \geq V_{0}\left(
c\right)  \text{.}%
\]
Typically, (see Duffie and Epstein \cite[p. 373]{DE} for example), dynamic
consistency is defined so as to deal also with strict rankings, that is, if
also $V_{t}\left(  c^{\prime}\right)  >V_{t}\left(  c\right)  $ on a
\textquotedblleft non-negligible\textquotedblright\ set of states, then
$V_{0}\left(  c^{\prime}\right)  >V_{0}\left(  c\right)  $. This added
requirement rules out the possibility that $c^{\prime}$ is chosen ex ante
though it is indifferent to $c$, and yet it is not implemented fully because
the individual switches to the conditionally strictly preferable $c$ for some
states at time $t$. The issue is how to specify \textquotedblleft
non-negligible\textquotedblright. When all priors are equivalent, then
positive probability according to any single prior is the natural
specification. In the absence of equivalence a similarly natural specification
is unclear. In particular, as illustrated in \cite{ej}, it is possible that
$c^{\prime}$ and $c$ be indifferent ex ante and yet that: (i) $c^{\prime}\geq
c$ on $\left[  0,T\right]  $ and they coincide on $\left[  0,t\right]  $; (ii)
there exists $t>0$, an event $N_{t}\in\mathcal{F}_{t}$, with $P\left(
N_{t}\right)  >0$ for some $P\in\mathcal{P}$, such that $c_{\tau}^{\prime
}>c_{\tau}$ for $t<\tau\leq T$ and $\omega\in N_{\tau}$. Then monotonicity of
preference would imply that $c^{\prime}$ be weakly preferable to $c$ at $t$
and strictly preferable conditionally on $\left(  t,N_{t}\right)  $, contrary
to the strict form of dynamic consistency.\footnote{The reason that
$c^{\prime}$ and $c$ are indifferent ex ante is that $N_{t}$ is ex ante null
according to the worst-case measure for $c$.} The fact that utility is
recursive but not strictly so suggests that, given an optimization problem,
though not every time $0$ optimal plan may be pursued subsequently at all
relevant nodes, under suitable regularity conditions there will exist at least
one time $0$ optimal plan that will be implemented. This is the case for an
example in \cite{ej}, but a general analysis remains to be done.

\newpage

\appendix

\section{Appendix: Uniform Continuity\label{app-uniform}}

Define the shifted canonical space by%
\[
\Omega^{t}\equiv\{ \omega\in C^{d}([t,T])\mid\omega_{t}=0\}.
\]
Denote by $B^{t}$ the canonical process on $\Omega^{t}$, (a shift of $B$), by
$P_{0}^{t}$ the probability measure on $\Omega^{t}$ such that $B^{t}$ is a
Brownian motion, and by $\mathcal{F}^{t}=\{ \mathcal{F}_{\tau}^{t}%
\}_{t\leq\tau\leq T}$ the filtration generated by $B^{t}$.

Fix $0\leq s\leq t\leq T.$ For $\omega\in\Omega^{s}$ and $\tilde{\omega}%
\in\Omega^{t}$, the concatenation of $\omega$ and $\tilde{\omega}$ at $t$ is
the path%
\[
(\omega\otimes_{t}\tilde{\omega})_{\tau}\triangleq\omega_{\tau}1_{[s,t)}%
(\tau)+(\omega_{t}+\tilde{\omega}_{\tau})1_{[t,T]}(\tau),\;s\leq\tau\leq T.
\]
Given an $\mathcal{F}_{T}^{s}$-measurable random variable $\xi$ on $\Omega
^{s}$ and $\omega\in\Omega^{s}$, define the shifted random variable
$\xi^{t,\omega}$ on $\Omega^{t}$ by%
\[
\xi^{t,\omega}(\tilde{\omega})\triangleq\xi(\omega\otimes_{t}\tilde{\omega
}),\; \tilde{\omega}\in\Omega^{t}.
\]
For an $\mathcal{F}^{s}$-progressively measurable process $(X_{\tau}%
)_{s\leq\tau\leq T}$, the shifted process $(X_{\tau}^{t,\omega})_{t\leq
\tau\leq T}$ is $\mathcal{F}^{t}$-progressively measurable.

Let $\left(  \Theta_{s}\right)  _{0\leq s\leq T}$ be a process of
correspondences as in Section \ref{section-priors}. For each $(t,\omega)$ in
$[0,T]\times\Omega$, define{\Large \ }a new process of correspondences
$(\Theta_{s}^{t,\omega})_{t\leq s\leq T}$ by:
\[
\Theta_{s}^{t,\omega}(\tilde{\omega})\triangleq\Theta_{s}(\omega\otimes
_{t}\tilde{\omega}),\; \tilde{\omega}\in\Omega^{t}.
\]
Then
\[
\Theta_{s}^{t,\omega}:\Omega^{t}\rightsquigarrow\mathbb{R}^{d}\times
\mathbb{R}^{d\times d}\text{.}%
\]
The new process inherits conditions (i)-(iv) and (vi). The same is true for
(v), which we define next.

The following definition is adapted from Nutz \cite[Defn. 3.2]{nutz}. Say that
$(\Theta_{t})$\ is \emph{uniformly continuous} if for all $\delta>0$ and
$(t,\omega)\in\lbrack0,T]\times\Omega$ there exists $\epsilon(t,\omega
,\delta)>0$ such that if $\sup_{0\leq s\leq t}\mid\omega_{s}-\omega
_{s}^{\prime}\mid\leq\epsilon$, then%
\[
ri^{\delta}\Theta_{s}^{t,\omega}(\tilde{\omega})\subseteq ri^{\epsilon}%
\Theta_{s}^{t,\omega^{\prime}}(\tilde{\omega})\; \text{for all }%
(s,\tilde{\omega})\in\lbrack t,T]\times\Omega^{t}.
\]

The process $\left(  \Theta_{t}\right)  $, and hence also $\Theta$, are fixed
throughout the appendices. Thus we write $\mathcal{P}$ instead of
$\mathcal{P}^{\Theta}$. Define $\mathcal{P}^{0}\subset\mathcal{P}$ by
\begin{equation}
\mathcal{P}^{0}\equiv\{P\in\mathcal{P}:\exists\delta>0\text{ }\theta
_{t}(\omega)\in ri^{\delta}\Theta_{t}(\omega)\; \text{for all }\left(
t,\omega\right)  \in\lbrack0,T]\times\Omega\}.\label{P0}%
\end{equation}

\section{Appendix: Conditioning\label{app-conditioning}}

Theorem \ref{thm-conditioning} is proven here.

In fact, we prove more than is stated in the theorem because we prove also the
following representation for conditional expectation. For each $t\in
\lbrack0,T]$\ and $P\in\mathcal{P}$, define
\begin{equation}
\mathcal{P}(t,P)=\{P^{\prime}\in\mathcal{P}:P^{\prime}=P\text{ on }%
\mathcal{F}_{t}\}.\label{PtP}%
\end{equation}
Then for each $\xi\in\widehat{L^{2}}(\Omega)$, $t\in\lbrack0,T]$ and
$P\in\mathcal{P}$,\footnote{Because all measures in $\mathcal{P}(t,P)$
coincide with $P$ on $\mathcal{F}_{t}$, essential supremum is defined as in
the classical case (see He et al. \cite[pp. 8-9]{He}, for example). Thus the
right hand side of (\ref{Econditional}) is defined to be any random variable
$\xi^{\ast}$ satisfying: (i) $\xi^{\ast}$ is $\mathcal{F}_{t}$-measurable,
$E_{P^{\prime}}[\xi\mid\mathcal{F}_{t}]\leq\xi^{\ast}$ $P$-$a.e.$ and (ii)
$\xi^{\ast}\leq\xi^{\ast\ast}$ $P$-$a.e.$ for any other random variable
$\xi^{\ast\ast}$ satisfying (i).}
\begin{equation}
\hat{E}[\xi\mid\mathcal{F}_{t}]=\underset{P^{\prime}\in\mathcal{P}%
(t,P)}{\text{ess}\sup}E_{P^{\prime}}[\xi\mid\mathcal{F}_{t}]\text{,
\ \ \ }P\text{-}a.e.\label{Econditional}%
\end{equation}

Perspective on this representation follows from considering the special case
where all measures in $\mathcal{P}$ are equivalent. Fix a measure $P_{0}$ in
$\mathcal{P}$. Then the condition (\ref{Econditional}) becomes%
\[
\hat{E}[\xi\mid\mathcal{F}_{t}]=~\underset{P^{\prime}\in\mathcal{P}%
(t,P)}{\text{ess}\sup}E_{P^{\prime}}[\xi\mid\mathcal{F}_{t}]\text{,
\ \ \ }P_{0}\text{-}a.e.\text{, for every }P\in\mathcal{P}.
\]
Accordingly, the random variable on the right side is (up to $P_{0}$-nullity)
independent of $P$. Apply $\cup_{P\in\mathcal{P}}\mathcal{P}(t,P)=\mathcal{P}$
to conclude that\footnote{The $P_{0}$-null event can be chosen independently
of $P$ by He et al. \cite[Theorem 1.3]{He}: Let $\mathcal{H}$ be a non-empty
family of random variables on any probability space. Then the essential
supremum exists and there is a countable number of elements $(\xi_{n})$ of
$\mathcal{H}$ such that ess$\sup\mathcal{H}=\underset{n}{\vee}\xi_{n}.$}
\[
\hat{E}[\xi\mid\mathcal{F}_{t}]=\text{ess}\sup_{P^{\prime}\in\mathcal{P}%
}E_{P^{\prime}}[\xi\mid\mathcal{F}_{t}]\text{, \ \ \ }P_{0}\text{-}a.s.
\]

\noindent In other words, conditioning amounts to applying the usual Bayesian
conditioning to each measure in $\mathcal{P}$ and taking the upper envelope of
the resulting expectations. This coincides with the prior-by-prior Bayesian
updating rule in the Chen-Epstein model (apart from the different convention
there of formulating expectations using infima rather than suprema). When the
set $\mathcal{P}$ is undominated, different measures in $\mathcal{P}$
typically provide different perspectives on any random variable. Accordingly,
(\ref{Econditional}) describes $\hat{E}[\xi\mid\mathcal{F}_{t}]$ completely by
describing how it appears when seen through the lens of every measure in
$\mathcal{P}$.

\bigskip

Our proof adapts arguments from Nutz \cite{nutz}. He constructs a time
consistent sublinear expectation in a setting with ambiguity about volatility
but not about drift. Because of this difference and because we use a different
approach to construct the set of priors, his results do not apply directly.

For any probability measure $P$ on the canonical space $\Omega$, a
corresponding \emph{regular conditional probability }$P_{t}^{\omega}$ is
defined to be any mapping $P_{t}^{\omega}:\Omega\times\mathcal{F}%
_{T}\rightarrow\lbrack0,1]$ satisfying the following conditions:

(i) for any $\omega$, \ $P_{t}^{\omega}$ is a probability measure on
$(\Omega,\mathcal{F}_{T})$.

(ii) for any $A\in F_{T}$, $\omega\rightarrow P_{t}^{\omega}(A)$\ is
$\mathcal{F}_{t}$-measurable.

(iii) for any $A\in\mathcal{F}_{T}$, $E^{P}[1_{A}\mid\mathcal{F}_{t}%
](\omega)=P_{t}^{\omega}(A),\;P$-$a.e.$

\noindent Of course, $P_{t}^{\omega}$\ is not defined uniquely by these
properties. We will fix a version defined via (\ref{Xrcp}) after proving in
Lemma \ref{lemma-rcp} that $P_{t}^{\omega}$ defined there satisfies the
conditions characterizing a regular conditional probability. This explains our
use of the same notation $P_{t}^{\omega}$ in both instances.

If $P$ is a probability on $\Omega^{s}$ and $\omega\in\Omega^{s}$, for any
$A\in\mathcal{F}_{T}^{t}$ we define%
\[
P^{t,\omega}(A)\triangleq P_{t}^{\omega}(\omega\otimes_{t}A),
\]
where $\omega\otimes_{t}A\triangleq\{ \omega\otimes_{t}\tilde{\omega}%
\mid\tilde{\omega}\in A\}$.

For each $(t,\omega)\in\lbrack0,T]\times\Omega$, let
\[
\theta_{s}(\tilde{\omega})=(\mu_{s}(\tilde{\omega}),\sigma_{s}(\tilde{\omega
}))\in ri^{\delta}\Theta_{s}^{t,\omega}(\tilde{\omega})\;\text{for all
}(s,\tilde{\omega})\in\lbrack t,T]\times\Omega^{t},
\]
where $\delta>0$ is some constant. Let $X^{t,\theta}=(X_{s}^{t,\theta})$ be
the solution of the following equation (under $P_{0}^{t}$)%
\[
dX_{s}^{t,\theta}=\mu_{s}(X_{\cdot}^{t,\theta})ds+\sigma_{s}(X_{\cdot
}^{t,\theta})dB_{s}^{t}\text{,}\;X_{t}^{t,\theta}=0\text{,}\;s\in\lbrack t,T].
\]
Then $X^{t,\theta}$ and $P_{0}^{t}$ induce a probability measure $P^{t,\theta
}$ on $\Omega^{t}$.

\begin{remark}
For nonspecialists we emphasize the difference between the preceding SDE and
(\ref{Xrcp}). The former is defined on the time interval $[t,T]$, and is a
shifted version of (\ref{Xtheta}), while (\ref{Xrcp}) is defined on the full
interval $[0,T]$. This difference is reflected also in the difference between
the induced measures: the shifted measure $P_{0}^{t}\in\Delta\left(
\Omega^{t}\right)  $ and the conditional measure $P_{t}^{\omega}\in
\Delta\left(  \Omega\right)  $. Part of the analysis to follow concerns
shifted SDE's, random variables and measures and their relation to unshifted
conditional counterparts. (See also Appendix \ref{app-uniform}.)
\end{remark}

Let $\mathcal{P}^{0}(t,\omega)$ be the collection of such induced measures
$P^{t,\theta}$. Define \newline$\deg(t,\omega,P^{t,\theta})=\delta^{\ast}%
/2>0$, where $\delta^{\ast}$ is the supremum of all $\delta$ such that%
\[
\theta_{s}(\tilde{\omega})\in ri^{\delta}\Theta_{s}^{t,\omega}(\tilde{\omega
})\; \text{for all }s\text{ and }\widetilde{\omega}\text{.}%
\]
Note that at time $t=0$, $\mathcal{P}^{0}(0,\omega)$ does not depend
on{\Large \ }$\omega${\Large \ }and it coincides with $\mathcal{P}^{0}$.

For each $(t,\omega)\in\lbrack0,T]\times\Omega$, let $\mathcal{P}(t,\omega)$
be the collection of all induced measures $P^{t,\theta}$ such that%
\[
\theta_{s}(\tilde{\omega})=(\mu_{s}(\tilde{\omega}),\sigma_{s}(\tilde{\omega
}))\in\Theta_{s}^{t,\omega}(\tilde{\omega})\; \text{for all }(s,\tilde{\omega
})\in\lbrack t,T]\times\Omega^{t}\text{.}%
\]
Note that $\mathcal{P}(0,\omega)=\mathcal{P}$.

\bigskip

Now we investigate the relationship between $\mathcal{P}(t,\omega)$ and
$\mathcal{P}_{t}^{\omega}$. (Recall that for any $P=P^{\theta}$ in
$\mathcal{P}$, the measure $P_{t}^{\omega}$ is defined via (\ref{Xrcp}) and
$\mathcal{P}_{t}^{\omega}$ is the set of all such measures as in
(\ref{regcond}).)

For any $\theta=(\mu,\sigma)\in\Theta$, $(t,\omega)\in\lbrack0,T]\times\Omega
$, define the shifted process $\bar{\theta}$ by%
\begin{equation}
\bar{\theta}_{s}(\tilde{\omega})=(\bar{\mu}_{s}(\tilde{\omega}),\bar{\sigma
}_{s}(\tilde{\omega}))\triangleq(\mu_{s}^{t,\omega}(\tilde{\omega}),\sigma
_{s}^{t,\omega}(\tilde{\omega}))\text{ \ for }\left(  s,\tilde{\omega}\right)
\in\lbrack t,T]\times\Omega^{t}\text{.}\label{Thetabar}%
\end{equation}
Then $\bar{\theta}_{s}(\tilde{\omega})\in\Theta_{s}^{t,\omega}(\tilde{\omega
})$. Consider the equation%
\begin{equation}
\left\{
\begin{array}
[c]{rl}%
d\bar{X}_{s} & =\bar{\mu}_{s}(\bar{X}.)ds+\bar{\sigma}_{s}(\bar{X}.)dB_{s}%
^{t}\text{,}\;s\in\lbrack t,T],\\
\bar{X}_{t} & =0.
\end{array}
\right. \label{Xbar}%
\end{equation}
Under $P_{0}^{t}$, the solution $\overline{X}$ induces a probability measure
$P^{t,\bar{\theta}}$ on $\Omega^{t}$. By the definition of $\mathcal{P}%
(t,\omega)$, $P^{t,\bar{\theta}}\in\mathcal{P}(t,\omega)$.

\begin{lemma}
\label{lemma-Ptw}$\{(P^{\prime})^{t,\omega}:P^{\prime}\in\mathcal{P}%
_{t}^{\omega}\}=\mathcal{P}(t,\omega)$.
\end{lemma}%

\textbf{Proof.}
$\subset$: For any $\theta=(\mu,\sigma)\in\Theta$, $P=P^{\theta}$ and
$(t,\omega)\in\lbrack0,T]\times\Omega$, we claim that
\[
(P_{t}^{\omega})^{t,\omega}=P^{t,\bar{\theta}}%
\]
where $P^{t,\bar{\theta}}$ is defined through (\ref{Xbar}). Because $B$ has
independent increments under $P_{0}$, the shifted solution $(X_{s}^{t,\omega
})_{t\leq s\leq T}$ of (\ref{Xrcp}) has the same distribution as does
$(\bar{X}_{s})_{t\leq s\leq T}$. This proves the claim.

$\supset$: Prove that for any $P^{t,\bar{\theta}}\in\mathcal{P}(t,\omega)$,
there exists $\theta\in\Theta$ such that (where $P=P^{\theta}$)
\[
P_{t}^{\omega}\in\mathcal{P}_{t}^{\omega}\; \text{and\ }(P_{t}^{\omega
})^{t,\omega}=P^{t,\bar{\theta}}.
\]
Each $P^{t,\bar{\theta}}$ in $\mathcal{P}(t,\omega)$ is induced by the
solution $\overline{X}$ of (\ref{Xbar}), where $\overline{\theta}\left(
\widetilde{\omega}\right)  $, defined in (\ref{Thetabar}), lies in $\Theta
_{s}^{t,\omega}(\tilde{\omega})$ for every $\tilde{\omega}\in\Omega^{t}$. For
any $\theta\in\Theta$ such that%
\[
\theta_{s}^{t,\omega}(\tilde{\omega})=(\mu_{s}^{t,\omega}(\tilde{\omega
}),\sigma_{s}^{t,\omega}(\tilde{\omega}))=(\bar{\mu}_{s}(\tilde{\omega}%
),\bar{\sigma}_{s}(\tilde{\omega}))\text{,}\;s\in\lbrack t,T],
\]
consider the equation (\ref{Xrcp}). Under $P_{0}$, the solution $X$ induces a
probability $P_{t}^{\omega}$ on $\Omega$. Because $B$ has independent
increments under $P_{0}$, we know that $(P_{t}^{\omega})^{t,\omega}%
=P^{t,\bar{\theta}}$. \ This completes the proof.\hfill$\blacksquare$

\begin{lemma}
\label{lemma-rcp}For any $\theta\in\Theta$ and $P=P^{\theta}$, $\{P_{t}%
^{\omega}:$ $(t,\omega)\in\lbrack0,T]\times\Omega\}$ is a version of the
regular conditional probability of $P$.
\end{lemma}%

\textbf{Proof.}
Firstly, for any $0<t_{1}<\cdots<t_{n}\leq T$ and bounded, continuous
functions $\varphi$ and $\psi$, we prove that%
\begin{equation}
E^{P}[\varphi(B_{t_{1}\wedge t},\ldots,B_{t_{n}\wedge t})\psi(B_{t_{1}}%
,\ldots,B_{t_{n}})]=E^{P}[\varphi(B_{t_{1}\wedge t},\ldots,B_{t_{n}\wedge
t})\psi_{t}]\label{rcp2}%
\end{equation}
where $t\in\lbrack t_{k},t_{k+1})$ and, for any $\hat{\omega}\in\Omega,$%
\[
\psi_{t}(\hat{\omega})\triangleq E^{P^{t,\bar{\theta}}}[\psi(\hat{\omega
}(t_{1}),\ldots,\hat{\omega}(t_{k}),\hat{\omega}(t)+B_{t_{k+1}}^{t}%
,\ldots,\hat{\omega}(t)+B_{t_{n}}^{t})].
\]

Note that $P^{t,\bar{\theta}}$ on $\Omega^{t}$ is induced by $\bar{X}$ (see
(\ref{Thetabar})) under $P_{0}^{t}$, and $P=P^{\theta}$ on $\Omega$ is induced
by $X=X^{\theta}$ (see (\ref{Xtheta})) under $P_{0}$. Then,%
\begin{align}
& \psi_{t}(X(\omega))=\label{rcp3}\\
& E^{P_{0}^{t}}[\psi(X_{t_{1}}(\omega),\ldots,X_{t_{k}}(\omega),X_{t}%
(\omega)+\bar{X}_{t_{k+1}}(\tilde{\omega}),\ldots,X_{t}(\omega)+\bar{X}%
_{t_{n}}(\tilde{\omega}))].\nonumber
\end{align}
Because $B$ has independent increments under $P_{0}$, the shifted regular
conditional probability%
\[
P_{0}^{t,\omega}=P_{0}^{t},\;P_{0}\text{-}a.e.
\]
Thus (\ref{rcp3}) holds under probability $P_{0}^{t,\omega}$.

Because $P_{0}^{t,\omega}$ is the shifted probability of $(P_{0})_{t}^{\omega
}$, we have%

\[%
\begin{array}
[c]{rl}
& \psi_{t}(X(\omega))\\
= & E^{(P_{0})_{t}^{\omega}}[\psi(X_{t_{1}}(\omega),\ldots,X_{t_{k}}%
(\omega),X_{t_{k+1}}(\omega),\ldots,X_{t_{n}}(\omega))]\\
= & E^{P_{0}}[\psi(X_{t_{1}}(\omega),\ldots,X_{t_{k}}(\omega),X_{t_{k+1}%
}(\omega),\ldots,X_{t_{n}}(\omega))\mid\mathcal{F}_{t}](\omega),\;P_{0}%
\text{-}a.e.
\end{array}
\]
Further, because $P$ is induced by $X$ and $P_{0}$,
\[%
\begin{array}
[c]{rl}
& E^{P}[\varphi(B_{t_{1}\wedge t},\ldots,B_{t_{n}\wedge t})\psi_{t}]\\
= & E^{P_{0}}[\varphi(X_{t_{1}\wedge t},\ldots,X_{t_{n}\wedge t})\psi
_{t}(X)]\\
= & E^{P_{0}}[\varphi(X_{t_{1}\wedge t},\ldots,X_{t_{n}\wedge t})E^{P_{0}%
}[\psi(X_{t_{1}},\ldots,X_{t_{n}})\mid\mathcal{F}_{t}]]\\
= & E^{P_{0}}[\varphi(X_{t_{1}\wedge t},\ldots,X_{t_{n}\wedge t})\psi
(X_{t_{1}},\ldots,X_{t_{n}})]\\
= & E^{P}[\varphi(B_{t_{1}\wedge t},\ldots,B_{t_{n}\wedge t})\psi(B_{t_{1}%
},\ldots,B_{t_{n}})].
\end{array}
\]

Secondly, note that (\ref{rcp2}) is true and $\varphi$ and $(t_{1}%
,\ldots,t_{n})$ are arbitrary. Then by the definition of the regular
conditional probability, for $P$-$a.e.\; \hat{\omega}\in\Omega$ and
$t\in\lbrack t_{k},t_{k+1})$,%
\begin{equation}
\psi_{t}(\hat{\omega})=E^{\tilde{P}^{t,\hat{\omega}}}[\psi(\hat{\omega}%
(t_{1}),\ldots,\hat{\omega}(t_{k}),\hat{\omega}(t)+B_{t_{k+1}}^{t},\ldots
,\hat{\omega}(t)+B_{t_{n}}^{t})]\text{,}\label{rcp4}%
\end{equation}
where $\tilde{P}^{t,\hat{\omega}}$ is the shift of the regular conditional
probability of $P$ given $(t,\hat{\omega})\in\lbrack0,T]\times\Omega.$

By standard approximating arguments, there exists a set $M$ such that $P(M)=0
$ and for any $\omega\notin M$, (\ref{rcp4}) holds for all continuous bounded
function $\psi$ and $(t_{1},\ldots,t_{n})$. This means that for $\omega\notin
M$ and for all bounded $\mathcal{F}_{T}^{t}$-measurable random variables $\xi$%
\[
E^{\tilde{P}^{t,\omega}}\xi=E^{P^{t,\bar{\theta}}}\xi.
\]
Then $\tilde{P}^{t,\omega}=P^{t,\bar{\theta}}$ $\ P$-$a.e.$ By Lemma
\ref{lemma-Ptw}, $\tilde{P}^{t,\omega}=P^{t,\bar{\theta}}=(P_{t}^{\omega
})^{t,\omega}$, $P$-$a.e.$ Thus $P_{t}^{\omega}$ is a version of the regular
conditional probability for $P$.\hfill$\blacksquare$

\bigskip

In the following, we always use $P_{t}^{\omega}$ defined by (\ref{Xrcp}) as
the fixed version of regular conditional probability for $P\in\mathcal{P}$.
Thus
\[
E^{P_{t}^{\omega}}\xi=E^{P}[\xi\mid\mathcal{F}_{t}](\omega)\text{, ~}%
P\text{-}a.e.\text{ }%
\]

Because we will want to consider also dynamics beginning at arbitrary $s$, let
$0\leq s\leq T$, $\bar{\omega}\in\Omega$, and $P\in\mathcal{P}(s,\bar{\omega
})$. Then given $(t,\omega)\in\lbrack s,T]\times\Omega^{s}$, we can fix a
version of the regular conditional probability, also denoted $P_{t}^{\omega}$
(here a measure on $\Omega^{s}$), which is constructed in a similar fashion
via a counterpart of (\ref{Xrcp}). Define
\[
\mathcal{P}_{t}^{\omega}(s,\bar{\omega})=\{P_{t}^{\omega}:P\in\mathcal{P}%
(s,\bar{\omega})\} \text{ and}%
\]%
\[
\mathcal{P}_{t}^{0,\omega}(s,\bar{\omega})=\{P_{t}^{\omega}:P\in
\mathcal{P}^{0}(s,\bar{\omega})\}.
\]
In each case, the obvious counterpart of the result in Lemma \ref{lemma-Ptw}
is valid.

\bigskip

\noindent

The remaining arguments are divided into three steps. First we prove that if
$\xi\in UC_{b}(\Omega)$ and if the set $\mathcal{P}$ is replaced by
$\mathcal{P}^{0}$ defined in (\ref{P0}), then the counterparts of
(\ref{Econditional}) and (\ref{LIE}) are valid. Then we show that
$\mathcal{P}^{0} $ (resp. $\mathcal{P}^{0}(t,\omega)$) is dense in
$\mathcal{P}$ (resp. $\mathcal{P}(t,\omega)$). In Step 3 the preceding is
extended to apply to all $\xi$\ in the completion of $UC_{b}(\Omega)$.

\bigskip

\noindent{\large Step 1}

Given $\xi\in UC_{b}(\Omega)$, define%
\begin{equation}
v_{t}^{0}(\omega)\triangleq\underset{P\in\mathcal{P}^{0}(t,\omega)}{\sup}%
E^{P}\xi^{t,\omega}\text{,}\; \ (t,\omega)\in\lbrack0,T]\times\Omega
.\label{vt0}%
\end{equation}

\begin{lemma}
\label{lemma-separability}Let $0\leq s\leq t\leq T$ and $\bar{\omega}\in
\Omega$. Given $\epsilon>0,$ there exist a sequence $(\hat{\omega}^{i}%
)_{i\geq1}$ in $\Omega^{s}$, an $\mathcal{F}_{t}^{s}$-measurable countable
partition $(E^{i})_{i\geq1}$ of $\Omega^{s}$, and a sequence $(P^{i})_{i\geq
1}$ of probability measures on $\Omega^{t}$ such that
\end{lemma}

(i) $\parallel\omega-\hat{\omega}^{i}\parallel_{\lbrack s,t]}\equiv\sup
_{s\leq\tau\leq t}\mid\omega_{\tau}-\hat{\omega}_{\tau}^{i}\mid\leq\epsilon$
for all $\omega\in E^{i};$

(ii) $P^{i}\in\mathcal{P}^{0}(t,\bar{\omega}\otimes_{s}\omega)$ for all
$\omega\in E^{i}$ and $\underset{\omega\in E^{i}}{\inf}$deg$(t,\bar{\omega
}\otimes_{s}\omega,P^{i})>0;$

(iii) $v_{t}^{0}(\bar{\omega}\otimes_{s}\hat{\omega}^{i})\leq E^{P^{i}}%
[\xi^{t,\bar{\omega}\otimes_{s}\hat{\omega}^{i}}]+\epsilon$.%

\textbf{Proof.}
Given $\epsilon>0$ and $\hat{\omega}\in\Omega^{s}$, by (\ref{vt0}) there
exists $P(\hat{\omega})\in\mathcal{P}^{0}(t,\bar{\omega}\otimes_{s}\hat
{\omega})$ such that
\[
v_{t}^{0}(\bar{\omega}\otimes_{s}\hat{\omega})\leq E^{P(\hat{\omega})}%
[\xi^{t,\bar{\omega}\otimes_{s}\hat{\omega}}]+\epsilon.
\]
Because $(\Theta_{t})$ is uniformly continuous, there exists $\epsilon
(\hat{\omega})>0$ such that
\[
P(\hat{\omega})\in\mathcal{P}^{0}(t,\bar{\omega}\otimes_{s}\omega^{\prime
})\text{ for all }\omega^{\prime}\in B(\epsilon(\hat{\omega}),\hat{\omega
})\text{ and }\underset{\omega^{\prime}\in B(\epsilon(\hat{\omega}%
),\hat{\omega})}{\inf}\text{deg}(t,\bar{\omega}\otimes_{s}\omega^{\prime
},P(\hat{\omega}))>0
\]
where $B(\epsilon,\hat{\omega})\triangleq\{ \omega^{\prime}\in\Omega^{s}%
\mid\parallel\omega^{\prime}-\hat{\omega}\parallel_{\lbrack s,t]}<\epsilon\}$
is the open $\parallel\cdot\parallel_{\lbrack s,t]}$ ball. Then $\{B(\epsilon
(\hat{\omega}),\hat{\omega})\mid\hat{\omega}\in\Omega^{s}\}$ forms an open
cover of $\Omega^{s}$. There exists a countable subcover because $\Omega^{s}$
is separable. We denote the subcover by%
\[
B^{i}\triangleq B(\epsilon(\hat{\omega}^{i}),\hat{\omega}^{i}),\;i=1,2,\ldots
\]
and define a partition of $\Omega^{s}$ by%
\[
E^{1}\triangleq B^{1},\;E^{i+1}\triangleq B^{i+1}\backslash(E^{1}\cup
\cdots\cup E^{i}),\;i\geq1.
\]

Set $P^{i}\triangleq P(\hat{\omega}^{i})$. Then (i)-(iii) are satisfied.\hfill
\hfill$\blacksquare$

\bigskip

For any $A\in\mathcal{F}_{T}^{s}$, define
\[
A^{t,\omega}=\{ \tilde{\omega}\in\Omega^{t}\mid\omega\otimes_{t}\tilde{\omega
}\in A\} \text{.}%
\]

\begin{lemma}
\label{lemma-pasting}Let $0\leq s\leq t\leq T$, $\bar{\omega}\in\Omega$ and
$P\in\mathcal{P}^{0}(s,\bar{\omega}).$ Let $(E^{i})_{0\leq i\leq N}$ be a
finite $\mathcal{F}_{t}^{s}$-measurable partition of $\Omega^{s}$. For $1\leq
i\leq N$, assume that $P^{i}\in\mathcal{P}^{0}(t,\bar{\omega}\otimes_{s}%
\omega)$ for all $\omega\in E^{i}$ and that $\underset{\omega\in E^{i}}{\inf}%
$deg$(t,\bar{\omega}\otimes_{s}\omega,P^{i})>0$. Define $\overline{P}$ by%
\[
\bar{P}(A)\triangleq P(A\cap E^{0})+%
{\displaystyle\sum\limits_{i=1}^{N}}
E^{P}[P^{i}(A^{t,\omega})1_{E^{i}}(\omega)],\ \;A\in\mathcal{F}_{T}%
^{s}\text{.}%
\]
Then: (i) $\bar{P}\in\mathcal{P}^{0}(s,\bar{\omega})$.

\ \ \ \ \ \ \ (ii) $\bar{P}=P$ on $\mathcal{F}_{t}^{s}$.

\ \ \ \ \ \ \ (iii) $\bar{P}^{t,\omega}=P^{t,\omega}$ $\ P$-$a.e.[\omega]$
\ on $E^{0}$.

\ \ \ \ \ \ \ (iv) $\bar{P}^{t,\omega}=P^{i}$ $\ \ ~P$-$a.e.[\omega]$ on
$E^{i}$, $1\leq i\leq N$.
\end{lemma}%

\textbf{Proof.}
(i) Let $\theta$ (resp. $\theta^{i})$ be the $\mathcal{F}^{s}$ (resp.
$\mathcal{F}^{t}$) -measurable process such that $P=P^{\theta}$ (resp.
$P^{i}=P^{\theta^{i}}).$ Define $\bar{\theta}$ by%
\begin{equation}
\bar{\theta}_{\tau}(\omega)\triangleq\theta_{\tau}(\omega)1_{[s,t)}%
(\tau)+[\theta_{\tau}(\omega)1_{E^{0}}(\omega)+%
{\displaystyle\sum\limits_{i=1}^{N}}
\theta_{\tau}^{i}(\omega^{t})1_{E^{i}}(\omega)]1_{[t,T]}(\tau)\label{paste}%
\end{equation}
for $(\tau,\omega)\in\lbrack s,T]\times\Omega^{s}$. Then $P^{\bar{\theta}}%
\in\mathcal{P}^{0}(s,\bar{\omega})$ and $\bar{P}=P^{\overline{\bar{\theta}}}$
on $\mathcal{F}_{T}^{s}$.

(ii) Let $A\in\mathcal{F}_{t}^{s}$. We prove $\bar{P}(A)=P(A)$. Note that for
$\omega\in\Omega^{s}$, if $\omega\in A$, then $A^{t,\omega}=\Omega^{t}$;
otherwise, $A^{t,\omega}=\emptyset$. Thus, $P^{i}(A^{t,\omega})=1_{A}(\omega)$
for $1\leq i\leq N$, and $\bar{P}(A)=P(A\cap E^{0})+%
{\displaystyle\sum\limits_{i=1}^{N}}
E^{P}[1_{A}(\omega)1_{E^{i}}(\omega)]=%
{\displaystyle\sum\limits_{i=0}^{N}}
E^{P}[A\cap E^{i}]=P(A)$.

(iii)-(iv) Recall the definition of $P_{t}^{\omega}$ by (\ref{Xrcp}). Note
that $\bar{P}=P^{\bar{\theta}}$ where $\bar{\theta}$ is defined by
(\ref{paste}). Then it is easy to show that the shifted regular conditional
probability $\bar{P}^{t,\omega}$ satisfies (iii)-(iv).\hfill$\blacksquare$

\bigskip

The technique used in proving Nutz \cite[Theorem 4.5]{nutz} can be adapted to
prove the following dynamic programming principle.

\begin{proposition}
\label{Proposition-dppo}Let $0\leq s\leq t\leq T$, $\xi\in UC_{b}(\Omega)$ and
define $v_{t}^{0}$ by (\ref{vt0}). Then
\begin{equation}
v_{s}^{0}(\bar{\omega})=\underset{P^{\prime}\in\mathcal{P}^{0}(s,\bar{\omega
})}{\sup}E^{P^{\prime}}[(v_{t}^{0})^{s,\bar{\omega}}]\; \text{for all }%
\bar{\omega}\in\Omega\text{,}\label{dppo1}%
\end{equation}%
\begin{equation}
v_{s}^{0}=\underset{P^{\prime}\in\mathcal{P}^{0}(s,P)}{\text{ess}\sup
}E^{P^{\prime}}[v_{t}^{0}\mid\mathcal{F}_{s}]\; \ P\text{-}a.e.\text{ for all
}P\in\mathcal{P}^{0}\text{,}\label{dppo2}%
\end{equation}
and%
\begin{equation}
v_{s}^{0}=\underset{P^{\prime}\in\mathcal{P}^{0}(s,P)}{\text{ess}\sup
}E^{P^{\prime}}[\xi\mid\mathcal{F}_{s}]\; \ \ P\text{-}a.e.\text{ for all
}P\in\mathcal{P}^{0}.\label{dppo3}%
\end{equation}

\end{proposition}%

\textbf{Proof.}
\textit{Proof of (\ref{dppo1})}: First prove $\leq$. Let $\bar{\omega}%
\in\Omega$ and $\omega\in\Omega^{s}$, by Lemma \ref{lemma-Ptw},%
\[
\{(P^{\prime})^{t,\omega}\mid P^{\prime}\in\mathcal{P}_{t}^{0,\omega}%
(s,\bar{\omega})\}=\mathcal{P}^{0}(t,\bar{\omega}\otimes_{s}\omega).
\]
For $P\in\mathcal{P}^{0}(s,\bar{\omega})$,%

\begin{equation}%
\begin{array}
[c]{rl}
& E^{P^{t,\omega}}[(\xi^{s,\bar{\omega}})^{t,\omega}]=E^{P^{t,\omega}}%
[\xi^{t,\bar{\omega}\otimes_{s}\omega}]\\
\leq & \underset{P^{\prime}\in\mathcal{P}^{0}(t,\bar{\omega}\otimes_{s}%
\omega)}{\sup}E^{P^{\prime}}[\xi^{t,\bar{\omega}\otimes_{s}\omega}]\\
= & v_{t}^{0}(\bar{\omega}\otimes_{s}\omega).
\end{array}
\label{dppo4}%
\end{equation}
Note that $v_{t}^{0}(\bar{\omega}\otimes_{s}\omega)=(v_{t}^{0})^{s,\bar
{\omega}}(\omega)$. Taking the expectation under $P$ on both sides of
(\ref{dppo4}) yields%
\[
E^{P}\xi^{s,\bar{\omega}}\leq E^{P}[(v_{t}^{0})^{s,\bar{\omega}}].
\]
The desired result holds by taking the supremum over $P\in\mathcal{P}%
^{0}(s,\bar{\omega})$.

Prove $\geq$. Let $\delta>0$. Because $^{t}\Omega$ is a Polish space and
$(v_{t}^{0})^{s,\bar{\omega}}$ is $\mathcal{F}_{t}^{s}$-measurable, by Lusin's
Theorem there exists a compact set $G\in\mathcal{F}_{t}^{s}$ with
$P(G)>1-\delta$ and such that $(v_{t}^{0})^{s,\bar{\omega}}$ is uniformly
continuous on $G$.

Let $\epsilon>0$. By Lemma \ref{lemma-separability}, there exist a sequence
$(\hat{\omega}^{i})_{i\geq1}$ in $G$, an $\mathcal{F}_{t}^{s}$-measurable
partition $(E^{i})_{i\geq1}$ of $G$, and a sequence $(P^{i})_{i\geq1}$ of
probability measures such that

(a) $\parallel\omega-\hat{\omega}^{i}\parallel_{\lbrack s,t]}\leq\epsilon$ for
all $\omega\in E^{i}$;

(b) $P^{i}\in\mathcal{P}^{0}(t,\bar{\omega}\otimes_{s}\omega)$ for all
$\omega\in E^{i}$ and $\underset{\omega\in E^{i}}{\inf}$deg$(t,\bar{\omega
}\otimes_{s}\omega,P^{i})>0$;

(c) $v_{t}^{0}(\bar{\omega}\otimes_{s}\hat{\omega}^{i})\leq E^{P^{i}}%
[\xi^{t,\bar{\omega}\otimes_{s}\hat{\omega}^{i}}]+\epsilon$.

\noindent Let
\[
A_{N}\triangleq E^{1}\cup\cdots\cup E^{N},\;N\geq1.
\]
For $P\in\mathcal{P}^{0}(s,\bar{\omega})$, define%
\[
\mathcal{P}^{0}(s,\bar{\omega},t,P)\triangleq\{P^{\prime}\in\mathcal{P}%
^{0}(s,\bar{\omega}):P^{\prime}=P\text{\ on }\mathcal{F}_{t}^{s}\} \text{.}%
\]
Apply Lemma \ref{lemma-pasting} to the finite partition $\{E^{1},\ldots
,E^{N},A_{N}^{c}\}$ of $\Omega^{s}$ to obtain a measure $P_{N}\in
\mathcal{P}^{0}(s,\bar{\omega})$ such that $P_{N}\in\mathcal{P}^{0}%
(s,\bar{\omega},t,P)$ and%
\begin{equation}
P_{N}^{t,\omega}=\left\{
\begin{array}
[c]{l}%
P^{t,\omega}\text{ for }\omega\in A_{N}^{c},\\
P^{i}\text{ for }\omega\in E^{i},\;1\leq i\leq N
\end{array}
\right. \label{PN}%
\end{equation}

Because $(v_{t}^{0})^{s,\bar{\omega}}$ and $\xi$ are uniformly continuous on
$G$, there exist moduli of continuity $\rho^{((v_{t}^{0})^{s,\bar{\omega}}\mid
G)}\left(  \cdot\right)  $ and $\rho^{(\xi)}\left(  \cdot\right)  $ such that%
\begin{align*}
& \mid(v_{t}^{0})^{s,\bar{\omega}}(\omega)-(v_{t}^{0})^{s,\bar{\omega}}%
(\omega^{\prime})\mid\leq\rho^{((v_{t}^{0})^{s,\bar{\omega}}\mid G)}%
(\parallel\omega-\omega^{\prime}\parallel_{\lbrack s,t]}),\\
& \mid\xi^{t,\bar{\omega}\otimes_{s}\omega}-\xi^{t,\bar{\omega}\otimes
_{s}\omega^{\prime}}\mid\leq\rho^{(\xi)}(\parallel\omega-\omega^{\prime
}\parallel_{\lbrack s,t]}).
\end{align*}

Let $\omega\in E^{i}$ for some $1\leq i\leq N$. Then%
\begin{equation}%
\begin{array}
[c]{rl}
& (v_{t}^{0})^{s,\bar{\omega}}(\omega)\\
\leq & (v_{t}^{0})^{s,\bar{\omega}}(\hat{\omega}^{i})+\rho^{((v_{t}%
^{0})^{s,\bar{\omega}}\mid G)}(\epsilon)\\
\leq & E^{P^{i}}[\xi^{t,\bar{\omega}\otimes_{s}\hat{\omega}^{i}}%
]+\epsilon+\rho^{((v_{t}^{0})^{s,\bar{\omega}}\mid G)}(\epsilon)\\
\leq & E^{P^{i}}[\xi^{t,\bar{\omega}\otimes_{s}\omega}]+\rho^{(\xi)}%
(\epsilon)+\epsilon+\rho^{((v_{t}^{0})^{s,\bar{\omega}}\mid G)}(\epsilon)\\
= & E^{P_{N}^{t,\omega}}[\xi^{t,\bar{\omega}\otimes_{s}\omega}]+\rho^{(\xi
)}(\epsilon)+\epsilon+\rho^{((v_{t}^{0})^{s,\bar{\omega}}\mid G)}(\epsilon)\\
= & E^{P_{N}}[\xi^{s,\bar{\omega}}\mid\mathcal{F}_{t}^{s}](\omega)+\rho
^{(\xi)}(\epsilon)+\epsilon+\rho^{((v_{t}^{0})^{s,\bar{\omega}}\mid
G)}(\epsilon)\; \text{for }P\text{-}a.e.\text{ }\omega\in E^{i}.
\end{array}
\label{dppo5}%
\end{equation}
These inequalities are due respectively to uniform continuity of $v_{t}^{0}$,
Lemma \ref{lemma-separability}(iii), uniform continuity of $\xi$, equation
(\ref{PN}), and the fact that $P_{N}\in\mathcal{P}(t,P)$. Because $P=P_{N}$ on
$\mathcal{F}_{t}^{s}$, taking the $P$-expectation on both sides yields%
\[
E^{P}[(v_{t}^{0})^{s,\bar{\omega}}1_{A_{N}}]\leq E^{P_{N}}[\xi^{s,\bar{\omega
}}1_{A_{N}}]+\rho^{(\xi)}(\epsilon)+\epsilon+\rho^{((v_{t}^{0})^{s,\bar
{\omega}}\mid G)}(\epsilon).
\]

Note that $\xi\in UC_{b}(\Omega)$ and $P_{N}(G\backslash A_{N})=P(G\backslash
A_{N})\rightarrow0$ as $N\rightarrow\infty$. Let $N\rightarrow\infty$ and
$\epsilon\rightarrow0$ to obtain that%
\[
E^{P}[(v_{t}^{0})^{s,\bar{\omega}}1_{G}]\leq\underset{P^{\prime}\in
\mathcal{P}^{0}(s,\bar{\omega},t,P)}{\sup}E^{P^{\prime}}[\xi^{s,\bar{\omega}%
}1_{G}].
\]
Because $\delta>0$ is arbitrary, similar arguments show that
\[
E^{P}[(v_{t}^{0})^{s,\bar{\omega}}]\leq\underset{P^{\prime}\in\mathcal{P}%
^{0}(s,\bar{\omega},t,P)}{\sup}E^{P^{\prime}}\xi^{s,\bar{\omega}}%
\leq\underset{P^{\prime}\in\mathcal{P}^{0}(s,\bar{\omega})}{\sup}E^{P^{\prime
}}\xi^{s,\bar{\omega}}=v_{s}^{0}(\bar{\omega}).
\]
But $P\in\mathcal{P}^{0}(s,\bar{\omega})$ is arbitrary. This completes the
proof of (\ref{dppo1}).

\bigskip

\noindent\textit{Proof of (\ref{dppo2})}: Fix $P\in\mathcal{P}^{0}$. First we
prove that
\begin{equation}
v_{t}^{0}\leq\underset{P^{\prime}\in\mathcal{P}^{0}(t,P)}{\text{ess}\sup
}E^{P^{\prime}}[\xi\mid\mathcal{F}_{t}]\; \ \ P\text{-}a.e.\label{dppo6}%
\end{equation}
Argue as in the second part of the preceding proof, specialized to $s=0$.
Conclude that there exists $P_{N}\in\mathcal{P}^{0}(t,P)$ such that, as a
counterpart of (\ref{dppo5}),
\[
v_{t}^{0}(\omega)\leq E^{P_{N}}[\xi\mid\mathcal{F}_{t}](\omega)+\rho^{(\xi
)}(\epsilon)+\epsilon+\rho^{(v_{t}^{0}\mid G)}(\epsilon)\; \text{for
}P\text{-}a.e.\text{ }\omega\in A_{N}.
\]
Because $P=P_{N}$ on $\mathcal{F}_{t}$, \ as $N\rightarrow\infty$ and
$\delta\rightarrow0$, one obtains (\ref{dppo6}).

Now prove the inequality $\leq$ in (\ref{dppo2}). For any $P^{\prime}%
\in\mathcal{P}^{0}(s,P)$, we know that $(P^{\prime})^{t,\omega}\in
\mathcal{P}^{0}(t,\omega)$. From (\ref{dppo1}),
\[
v_{t}^{0}(\omega)\geq E^{(P^{\prime})^{t,\omega}}[\xi^{t,\omega}%
]=E^{P^{\prime}}[\xi\mid\mathcal{F}_{t}](\omega)\;P^{\prime}\text{-}a.e.
\]
Taking the conditional expectation on both sides yields $E^{P^{\prime}}%
[\xi\mid\mathcal{F}_{s}]\leq E^{P^{\prime}}[v_{t}^{0}\mid\mathcal{F}%
_{s}]\;P^{\prime}$-$a.e.$, hence also $P$-$a.e.$ Thus
\[
v_{s}^{0}\leq\underset{P^{\prime}\in\mathcal{P}^{0}(s,P)}{\text{ess}\sup
}E^{P^{\prime}}[\xi\mid\mathcal{F}_{s}]\leq\underset{P^{\prime}\in
\mathcal{P}^{0}(s,P)}{\text{ess}\sup}E^{P^{\prime}}[v_{t}^{0}\mid
\mathcal{F}_{s}]\;P\text{-}a.e.
\]

Thirdly, we prove the converse direction holds in (\ref{dppo2}). For any
$P^{\prime}\in\mathcal{P}^{0}(s,P),$ by (\ref{dppo1}) we have%
\[
v_{s}^{0}(\omega)\geq E^{(P^{\prime})^{s,\omega}}[(v_{t}^{0})^{s,\omega
}]=E^{P^{\prime}}[v_{t}^{0}\mid\mathcal{F}_{s}](\omega)
\]
$P^{\prime}$-$a.e.$ on $\mathcal{F}_{s}$ and hence $P$-$a.e.$

Equation (\ref{dppo3}) is implied by (\ref{dppo2}) because $v_{T}^{0}=\xi
$.\hfill$\blacksquare$

\vspace{0.39in}

\noindent{\large STEP 2}

Refer to the topology induced on $\Delta\left(  \Omega\right)  $ by bounded
continuous functions as the weak-convergence topology.

\begin{lemma}
\label{lemma-dense}(a) $\mathcal{P}^{0}$ is dense in $\mathcal{P}$ in the weak
convergence topology.

(b) For each $t$ and $\omega$, $\mathcal{P}^{0}(t,\omega)$ is dense in
$\mathcal{P}(t,\omega)$ in the weak convergence topology.
\end{lemma}%

\textbf{Proof.}
(a) Let $P^{\theta^{0}}\in\mathcal{P}^{0}$ and $P^{\theta}\in\mathcal{P}$, and
define
\[
\theta^{\epsilon}=\epsilon\theta^{0}+(1-\epsilon)\theta\,,
\]
where $0<\epsilon<1$. By Uniform Interiority for $\left(  \Theta_{t}\right)
$, there exists $\delta>0\,$\ such that $\theta_{t}^{\epsilon}(\omega)\in
ri^{\delta}\Theta_{t}(\omega)$ for all $t$ and $\omega$. Thus $P^{\theta
^{\epsilon}}\in\mathcal{P}^{0}$.

By the standard approximation of a stochastic differential equation (see
Gihman and Skorohod \cite[Thm 3.15]{GSb}), as $\epsilon\rightarrow0$ there
exists a subsequence of $X^{\theta^{\epsilon}}$, which we still denote by
$X^{\theta^{\epsilon}}$, such that%
\[
\sup_{0\leq t\leq T}\mid X_{t}^{\theta^{\epsilon}}-X_{t}^{\theta}%
\mid\rightarrow0\;P_{0}\text{-}a.e.
\]
The Dominated Convergence Theorem implies that $P^{\theta^{\epsilon}%
}\rightarrow P^{\theta}$.

(b) The proof is similar.\hfill$\blacksquare$

\bigskip

For any $\theta=(\mu,\sigma)\in\Theta$, $(t,\omega)\in\lbrack0,T]\times\Omega
$, define the shifted process $\bar{\theta}$, the process $\overline{X}$ and
the probability measure $P^{t,\overline{\theta}}$ exactly as in
(\ref{Thetabar}) and (\ref{Xbar}). As noted earlier, $P^{t,\bar{\theta}}%
\in\mathcal{P}(t,\omega)$.

Recall that for any $P$ in $\mathcal{P}$, the measure $P_{t}^{\omega}$ is
defined via (\ref{Xrcp}); $\mathcal{P}_{t}^{\omega}$ is the set of all such
measures (\ref{regcond}). By construction,
\[
P_{t}^{\omega}\left(  \{ \bar{\omega}\in\Omega:\bar{\omega}_{s}=\omega
_{s}\;s\in\lbrack0,t]\} \right)  =1\text{.}%
\]
We show shortly that $P_{t}^{\omega}$ is a version of the regular conditional
probability for $P$.

Given $\xi\in UC_{b}(\Omega)$, define%
\begin{equation}
v_{t}(\omega)\triangleq\underset{P\in\mathcal{P}(t,\omega)}{\sup}E^{P}%
\xi^{t,\omega}\text{,}\;(t,\omega)\in\lbrack0,T]\times\Omega.\label{vt}%
\end{equation}

\begin{lemma}
\label{lemma-Econd}For any $(t,\omega)\in\lbrack0,T]\times\Omega$ and $\xi\in
UC_{b}(\Omega)$, we have
\begin{equation}
v_{t}(\omega)=v_{t}^{0}(\omega),\label{vtvt0}%
\end{equation}%
\begin{equation}
v_{t}(\omega)=\underset{P\in\mathcal{P}_{t}^{\omega}}{\sup}E^{P}\xi
\text{,}\label{vtrcp}%
\end{equation}
and%
\begin{equation}
v_{t}=\underset{P^{\prime}\in\mathcal{P}(t,P)}{\text{ess}\sup}E^{P^{\prime}%
}[\xi\mid\mathcal{F}_{t}]\;P\text{-}a.e.\text{ for all }P\in\mathcal{P}%
.\label{vtPtP}%
\end{equation}
Furthermore, for any{\Large \ }$0\leq s\leq t\leq T$,
\begin{equation}
v_{s}(\omega)=\underset{P^{\prime}\in\mathcal{P}\left(  s,\omega\right)
}{\sup}E^{P^{\prime}}[(v_{t})^{s,\omega}]\; \text{for all }\omega\in
\Omega\text{,}\label{vtvtc}%
\end{equation}
and%
\begin{equation}
v_{s}=\underset{P^{\prime}\in\mathcal{P}(s,P)}{\text{ess}\sup}E^{P^{\prime}%
}[v_{t}\mid\mathcal{F}_{s}]\; \ P\text{-}a.e.\text{ for all }P\in
\mathcal{P}\text{.}\label{vtPtc}%
\end{equation}

\end{lemma}%

\textbf{Proof.}
\textit{Proof of (\ref{vtvt0})}: It is implied by the fact that $\mathcal{P}%
^{0}(t,\omega)$ is dense in $\mathcal{P}(t,\omega)$.

\medskip

\noindent\textit{Proof of (\ref{vtrcp})}: By the definition of $v_{t}(\omega
)$, we know that%
\[
v_{t}(\omega)=\underset{P\in\mathcal{P}(t,\omega)}{\sup}E^{P}\xi^{t,\omega}.
\]
By Lemma \ref{lemma-Ptw},%
\[
\{(P^{\prime})^{t,\omega}\mid P^{\prime}\in\mathcal{P}_{t}^{\omega
}\}=\mathcal{P}(t,\omega).
\]
Thus
\[
v_{t}(\omega)=\underset{\hat{P}\in\mathcal{P}_{t}^{\omega}}{\sup}E^{\hat
{P}^{t,\omega}}\xi^{t,\omega}=\underset{\hat{P}\in\mathcal{P}_{t}^{\omega
}}{\sup}E^{\hat{P}}\xi.
\]

\noindent\textit{Proof of (\ref{vtPtP})}: Fix $P\in\mathcal{P}$. For any
$P^{\prime}\in\mathcal{P}(t,P)$, by Lemmas \ref{lemma-Ptw} and \ref{lemma-rcp}%
, $(P^{\prime})^{t,\omega}\in\mathcal{P}(t,\omega)$. By the definition of
$v_{t}(\omega)$,
\[
v_{t}(\omega)\geq E^{(P^{\prime})^{t,\omega}}\xi^{t,\omega}=E^{P^{\prime}}%
[\xi\mid\mathcal{F}_{t}](\omega)
\]
$P^{\prime}$-$a.e.$ on $\mathcal{F}_{t}$ and hence $P$-$a.e.$ Thus%

\[
v_{t}\geq\underset{P^{\prime}\in\mathcal{P}(t,P)}{\text{ess}\sup}E^{P^{\prime
}}[\xi\mid\mathcal{F}_{t}]\;P\text{-}a.e.
\]

Now we prove the reverse inequality. By (\ref{vtvt0}), $v_{t}(\omega
)=v_{t}^{0}(\omega)$. Then, using the same technique as in the proof of
Proposition \ref{Proposition-dppo} for the special case $s=0$, there exists
$P_{N}\in\mathcal{P}(t,P)$ such that, as a counterpart of (\ref{dppo5}),%
\[%
\begin{array}
[c]{rl}%
v_{t}(\omega) & \leq E^{P_{N}}[\xi\mid\mathcal{F}_{t}](\omega)+\rho^{(\xi
)}(\epsilon)+\epsilon+\rho^{(v_{t}\mid G)}(\epsilon)\\
& \leq\underset{N}{\sup}E^{P_{N}}[\xi\mid\mathcal{F}_{t}](\omega)+\rho^{(\xi
)}(\epsilon)+\epsilon+\rho^{(v_{t}\mid G)}(\epsilon)
\end{array}
\]
for $P$-$a.e.$ $\omega\in A_{N}$. \ Let $N\rightarrow\infty$ to obtain that,
for $P$-$a.e.$ $\omega\in G$,
\[
v_{t}(\omega)\leq\underset{P^{\prime}\in\mathcal{P}(t,P)}{\text{ess}\sup
}E^{P^{\prime}}[\xi\mid\mathcal{F}_{t}](\omega)+\rho^{(\xi)}(\epsilon
)+\epsilon+\rho^{(v_{t}\mid G)}(\epsilon)\text{.}%
\]
Let $\epsilon\rightarrow0$ to derive
\[
v_{t}\leq\underset{P^{\prime}\in\mathcal{P}(t,P)}{\text{ess}\sup}E^{P^{\prime
}}[\xi\mid\mathcal{F}_{t}]\text{,}\;P\text{-}a.e.\text{ on }G\text{.}%
\]
Note that $G$ depends on $\delta$, but not on $\epsilon$. Let $\delta
\rightarrow0$ and conclude that%
\[
v_{t}\leq\underset{P^{\prime}\in\mathcal{P}(t,P)}{\text{ess}\sup}E^{P^{\prime
}}[\xi\mid\mathcal{F}_{t}],\;P\text{-}a.e.
\]

\medskip

{\Large \noindent}\noindent\noindent

\noindent\textit{Proof of (\ref{vtvtc}) and (\ref{vtPtc}):} The former is due
to (\ref{dppo1}) and the fact that $P^{0}(t,\omega)$\ is dense in
$P(t,\omega)$. The proof of (\ref{vtPtc}) is similar to the proof of
(\ref{dppo3}) in Proposition \ref{Proposition-dppo}.\hfill$\blacksquare$

\vspace{0.4in}

Now for any $\xi\in UC_{b}(\Omega)$, we define conditional expectation by
\[
\hat{E}[\xi\mid\mathcal{F}_{t}](\omega)\triangleq v_{t}(\omega)\text{.}%
\]

\bigskip\vspace{0.4in}

\noindent{\Large STEP 3}

Thus far we have defined $\hat{E}[\xi\mid\mathcal{F}_{t}](\omega)$, for all
$t$ and $\omega$, for any $\xi\in UC_{b}(\Omega)$. Now extend the operator
$\hat{E}[\cdot\mid\mathcal{F}_{t}]$ to the completion of $UC_{b}(\Omega)$.

\begin{lemma}
[Extension]\label{lemma-extension}The mapping $\hat{E}[\cdot\mid
\mathcal{F}_{t}]$ on $UC_{b}(\Omega)$ can be extended uniquely to a
$1$-Lipschitz continuous mapping $\hat{E}[\cdot\mid\mathcal{F}_{t}%
]:\widehat{L^{2}}(\Omega)\rightarrow\widehat{L^{2}}(^{t}\Omega)$.
\end{lemma}%

\textbf{Proof.}
Define $\widetilde{L^{2}}(^{t}\Omega)$ to be the space of $\mathcal{F}_{t}%
$-measurable random variables $X$ satisfying%
\[
\parallel X\parallel\triangleq(\hat{E}[\mid X\mid^{2}])^{\frac{1}{2}}%
=(\sup_{P\in\mathcal{P}}E^{P}[\mid X\mid^{2}])^{\frac{1}{2}}<\infty.
\]
Obviously, $\widehat{L^{2}}(^{t}\Omega)\subset\widetilde{L^{2}}(^{t}\Omega)$.

(i) We prove that $\hat{E}[\cdot\mid\mathcal{F}_{t}]$ can be uniquely extended
to a $1$-Lipschitz continuous mapping%
\[
\hat{E}[\cdot\mid\mathcal{F}_{t}]:\widehat{L^{2}}(\Omega)\rightarrow
\widetilde{L^{2}}(^{t}\Omega).
\]
For any $\xi$ and $\eta$ in $UC_{b}(\Omega)$,
\[
\mid\hat{E}[\xi^{\prime}\mid\mathcal{F}_{t}]-\hat{E}[\xi\mid\mathcal{F}%
_{t}]\mid^{2}\leq\left(  \hat{E}[\left(  \mid\xi^{\prime}-\xi\mid\right)
\mid\mathcal{F}_{t}]\right)  ^{2}\leq\hat{E}[\mid\xi^{\prime}-\xi\mid^{2}%
\mid\mathcal{F}_{t}],
\]
where the first inequality follows primarily from the subadditivity of
$\hat{E}[\cdot\mid\mathcal{F}_{t}]$, and the second is implied by Jensen's
inequality applied to each $P$ in $\mathcal{P}$. Thus%
\begin{align*}
& \parallel\hat{E}[\xi^{\prime}\mid\mathcal{F}_{t}]-\hat{E}[\xi\mid
\mathcal{F}_{t}]\parallel\\
& =\left(  \widehat{E}\left[  \mid\hat{E}[\xi^{\prime}\mid\mathcal{F}%
_{t}]-\hat{E}[\xi\mid\mathcal{F}_{t}]\mid^{2}\right]  \right)  ^{1/2}\\
& \leq\left(  \widehat{E}\left[  \hat{E}[\mid\xi^{\prime}-\xi\mid^{2}%
\mid\mathcal{F}_{t}]\right]  \right)  ^{1/2}\\
& =\left(  \widehat{E}\left[  \mid\xi^{\prime}-\xi\mid^{2}\right]  \right)
^{1/2}=\parallel\xi^{\prime}-\xi\parallel\text{,}%
\end{align*}
where the second equality is due to the `law of iterated expectations' for
integrands in $UC_{b}(\Omega)$ proven in Lemma \ref{lemma-Econd}.

As a consequence, $\hat{E}[\xi\mid\mathcal{F}_{t}]\in\widetilde{L^{2}}%
(^{t}\Omega)$ \ for $\xi$ and $\eta$ in $UC_{b}(\Omega)$, and $\hat{E}%
[\cdot\mid\mathcal{F}_{t}]$ extends uniquely to a $1$-Lipschitz continuous
mapping from $\widehat{L^{2}}(\Omega)$ into $\widetilde{L^{2}}(^{t}\Omega)$.

\medskip

\noindent(ii) Now prove that $\hat{E}[\cdot\mid\mathcal{F}_{t}]$ maps
$\widehat{L^{2}}(\Omega)$ into $\widehat{L^{2}}(^{t}\Omega)$.

\noindent First we show that if $\xi\in UC_{b}(\Omega)$, then $\hat{E}[\xi
\mid\mathcal{F}_{t}]$ is lower semicontinuous: Fix $\omega\in\Omega$. Since
$\xi\in UC_{b}(\Omega)$, there exists a modulus of continuity $\rho^{(\xi)}$
such that for all $\omega^{\prime}\in\Omega$ and $\tilde{\omega}\in\Omega^{t}$%
\begin{align*}
& \mid\xi(\omega)-\xi(\omega^{\prime})\mid\leq\rho^{(\xi)}(\parallel
\omega-\omega^{\prime}\parallel_{\lbrack0,T]})\text{, and}\\
& \mid\xi^{t,\omega}(\tilde{\omega})-\xi^{t,\omega^{\prime}}(\tilde{\omega
})\mid\leq\rho^{(\xi)}(\parallel\omega-\omega^{\prime}\parallel_{\lbrack
0,t]})\text{.}%
\end{align*}
Consider a sequence $(\omega_{n})$ such that $\parallel\omega-\omega
_{n}\parallel_{\lbrack0,t]}\rightarrow0$. For any $P\in\mathcal{P}(t,\omega)$,
by the Uniform Continuity assumption for $\left(  \Theta_{t}\right)  $, we
know that $P\in\mathcal{P}(t,\omega_{n})$ when $n$ is large enough. Thus%
\[%
\begin{array}
[c]{rl}
& \underset{n\rightarrow\infty}{\lim\inf}v_{t}(\omega_{n})\\
= & \underset{n\rightarrow\infty}{\lim\inf}\underset{P^{\prime}\in
\mathcal{P}(t,\omega_{n})}{\sup}E^{P^{\prime}}\xi^{t,\omega_{n}}\\
\geq & \underset{n\rightarrow\infty}{\lim\inf}\left(  \left(
\underset{P^{\prime}\in\mathcal{P}(t,\omega_{n})}{\sup}E^{P^{\prime}}%
\xi^{t,\omega}\right)  -\rho^{(\xi)}(\parallel\omega-\omega_{n}\parallel
_{\lbrack0,t]}\right) \\
= & \underset{P^{\prime}\in\mathcal{P}(t,\omega_{n})}{\sup}E^{P^{\prime}}%
\xi^{t,\omega}\geq E^{P}\xi^{t,\omega}\,.
\end{array}
\]
Because $P\in\mathcal{P}(t,\omega)$ is arbitrary, this proves that $\lim
\inf_{n\longrightarrow\infty}v_{t}(\omega_{n})\geq v_{t}(\omega)$, which is
the asserted lower semicontinuity.

Next prove that any bounded lower semicontinuous function $f$ on $^{t}\Omega$
is in $\widehat{L^{2}}(^{t}\Omega)$: Because $^{t}\Omega$ is Polish, there
exists a uniformly bounded sequence $f_{n}\in C_{b}(^{t}\Omega)$ such that
$f_{n}\uparrow f$ for all $\omega\in\Omega$. By Gihman and Skorohod
\cite[Theorem 3.10]{GSb}, $\mathcal{P}$ is relatively compact in the weak
convergence topology. Therefore, by Tietze's Extension Theorem (Mandelkern
\cite{Man}), $C_{b}(^{t}\Omega)\subset\widehat{L^{2}}(^{t}\Omega)$, and by
Denis et al. \cite[Theorem 12]{DHP}, $\sup_{P\in\mathcal{P}}E^{P}\left(  \mid
f-f_{n}\mid^{2}\right)  \rightarrow0$. Thus $f\in\widehat{L^{2}}(^{t}\Omega)$.

Combine these two results to deduce that $\hat{E}[\xi\mid\mathcal{F}_{t}]$
$\in\widehat{L^{2}}(^{t}\Omega)$ if $\xi\in UC_{b}(\Omega)$. From (i), $\{
\hat{E}[\xi\mid\mathcal{F}_{t}]:\xi\in\widehat{L^{2}}(\Omega)\}$ is contained
in the $\parallel\cdot\parallel$-closure of $\{ \hat{E}[\xi\mid\mathcal{F}%
_{t}]:\xi\in UC_{b}(\Omega)\}$. But $\{ \hat{E}[\xi\mid\mathcal{F}_{t}]:\xi\in
UC_{b}(\Omega)\}$ is contained in $\widehat{L^{2}}(^{t}\Omega)$, which is
complete under $\parallel\cdot\parallel$. This completes the proof.
\hfill$\blacksquare$

\bigskip

\noindent\textit{Proof of (\ref{Econditional})}: Fix $P\in\mathcal{P}$ and
$X\in\widehat{L^{2}}(\Omega)$. Given $\epsilon>0$, there exists $\overline
{X}\in UC_{b}(\Omega)$ such that
\[
\parallel\hat{E}[X\mid\mathcal{F}_{t}]-\hat{E}[\overline{X}\mid\mathcal{F}%
_{t}]\parallel\leq\parallel X-\overline{X}\parallel\leq\epsilon.
\]
For any $P^{\prime}\in\mathcal{P}(t,P)$,
\begin{equation}%
\begin{array}
[c]{rl}
& E^{P^{\prime}}[X\mid\mathcal{F}_{t}]-\hat{E}[X\mid\mathcal{F}_{t}]\\
= & E^{P^{\prime}}[X-\overline{X}\mid\mathcal{F}_{t}]+(E^{P^{\prime}%
}[\overline{X}\mid\mathcal{F}_{t}]-\hat{E}[\overline{X}\mid\mathcal{F}%
_{t}])+(\hat{E}[\overline{X}\mid\mathcal{F}_{t}]-\hat{E}[X\mid\mathcal{F}%
_{t}]).
\end{array}
\label{exttc0}%
\end{equation}

From Karatzas and Shreve \cite[Theorem A.3]{KS}, derive that there exists a
sequence $P_{n}\in\mathcal{P}(t,P)$ such that%
\begin{equation}
\underset{P^{\prime}\in\mathcal{P}(t,P)}{\text{ess}\sup}E^{P^{\prime}%
}[\overline{X}\mid\mathcal{F}_{t}]=\underset{n\rightarrow\infty}{\lim}%
E^{P_{n}}[\overline{X}\mid\mathcal{F}_{t}]\; \ P\text{-}a.e.\label{exttc1}%
\end{equation}
where $P$-$a.e.$ the sequence on the right is increasing in $n$. Then by Lemma
\ref{lemma-Econd},%
\begin{equation}
\hat{E}[\overline{X}\mid\mathcal{F}_{t}]=\underset{P^{\prime}\in
\mathcal{P}(t,P)}{\text{ess}\sup}E^{P^{\prime}}[\overline{X}\mid
\mathcal{F}_{t}]=\underset{n\rightarrow\infty}{\lim}E^{P_{n}}[\overline{X}%
\mid\mathcal{F}_{t}]\; \ P\text{-}a.e.\label{exttc2}%
\end{equation}

Denote $L^{2}(\Omega,\mathcal{F}_{T},P)$ by $L^{2}(P)$. Compute $L^{2}%
(P)$-norms on both sides of (\ref{exttc0}) to obtain, for every $n$,%
\[%
\begin{array}
[c]{rl}
& \parallel E^{P_{n}}[X\mid\mathcal{F}_{t}]-\hat{E}[X\mid\mathcal{F}%
_{t}]\parallel_{L^{2}(P)}\\
\leq & \parallel X-\overline{X}\parallel_{L^{2}(P_{n})}+\parallel E^{P_{n}%
}[\overline{X}\mid\mathcal{F}_{t}]-\hat{E}[\overline{X}\mid\mathcal{F}%
_{t}]\parallel_{L^{2}(P)}+\parallel\hat{E}[\overline{X}\mid\mathcal{F}%
_{t}]-\hat{E}[X\mid\mathcal{F}_{t}]\parallel_{L^{2}(P)}\\
\leq & \parallel E^{P_{n}}[\overline{X}\mid\mathcal{F}_{t}]-\hat{E}%
[\overline{X}\mid\mathcal{F}_{t}]\parallel_{L^{2}(P)}+2\epsilon
\end{array}
\]
By (\ref{exttc2}),
\[
\underset{n\rightarrow\infty}{\lim\sup}\parallel E^{P_{n}}[X\mid
\mathcal{F}_{t}]-\hat{E}[X\mid\mathcal{F}_{t}]\parallel\leq2\epsilon.
\]
Note that $\epsilon$ is arbitrary. Therefore, there exists a sequence $\hat
{P}_{n}\in\mathcal{P}(t,P)$ such that $E^{\hat{P}_{n}}[X\mid\mathcal{F}%
_{t}]\rightarrow\hat{E}[X\mid\mathcal{F}_{t}],$ $P$-$a.e.$, which implies that%
\begin{equation}
\hat{E}[X\mid\mathcal{F}_{t}]\leq\underset{P^{\prime}\in\mathcal{P}%
(t,P)}{\text{ess}\sup}E^{P^{\prime}}[X\mid\mathcal{F}_{t}].\label{exttc3}%
\end{equation}

As in (\ref{exttc1}), there exists a sequence $P_{n}^{\prime}\in
\mathcal{P}(t,P)$ such that%
\[
\underset{P^{\prime}\in\mathcal{P}(t,P)}{\text{ess}\sup}E^{P^{\prime}}%
[X\mid\mathcal{F}_{t}]=\underset{n\rightarrow\infty}{\lim}E^{P_{n}^{\prime}%
}[X\mid\mathcal{F}_{t}]\; \ P\text{-}a.e.
\]
with the sequence on the right being increasing in $n$ ($P$-$a.e.$). Set
\newline$A_{n}\triangleq\{E^{P_{n}^{\prime}}[X\mid\mathcal{F}_{t}]\geq\hat
{E}[X\mid\mathcal{F}_{t}]\}$. By (\ref{exttc3}), $P$-$a.e.$%
\[
0\leq(E^{P_{n}^{\prime}}[X\mid\mathcal{F}_{t}]-\hat{E}[X\mid\mathcal{F}%
_{t}])1_{A_{n}}\overset{n}{\nearrow}\underset{P^{\prime}\in\mathcal{P}%
(t,P)}{\text{ess}\sup}E^{P^{\prime}}[X\mid\mathcal{F}_{t}]-\hat{E}%
[X\mid\mathcal{F}_{t}]\text{.}\;
\]

By (\ref{exttc0}) and (\ref{exttc2}), $P$-$a.e.$
\[
E^{P_{n}^{\prime}}[X\mid\mathcal{F}_{t}]-\hat{E}[X\mid\mathcal{F}_{t}]\leq
E^{P_{n}^{\prime}}[X-\overline{X}\mid\mathcal{F}_{t}]+(\hat{E}[\overline
{X}\mid\mathcal{F}_{t}]-\hat{E}[X\mid\mathcal{F}_{t}])
\]
Take $L^{2}(P)$-norms to derive%
\[%
\begin{array}
[c]{rl}
& \parallel\underset{P^{\prime}\in\mathcal{P}(t,P)}{\text{ess}\sup
}E^{P^{\prime}}[X\mid\mathcal{F}_{t}]-\hat{E}[X\mid\mathcal{F}_{t}%
]\parallel_{L^{2}(P)}\\
= & \underset{n\rightarrow\infty}{\lim}\parallel(E^{P_{n}^{\prime}}%
[X\mid\mathcal{F}_{t}]-\hat{E}[X\mid\mathcal{F}_{t}])1_{A_{n}}\parallel
_{L^{2}(P)}\\
\leq & \underset{n\rightarrow\infty}{\lim\sup}\parallel X-\overline
{X}\parallel_{L^{2}(P_{n}^{\prime})}+\parallel\hat{E}[\overline{X}%
\mid\mathcal{F}_{t}]-\hat{E}[X\mid\mathcal{F}_{t}]\parallel_{L^{2}(P)}\\
\leq & 2\epsilon.
\end{array}
\]
This proves (\ref{Econditional}).

\bigskip

\noindent\textit{Proof of (\ref{LIE})}: It is sufficient to prove that, for
$0\leq s\leq t\leq T$, $P$-$a.e.$%
\begin{equation}%
\begin{array}
[c]{rl}
& \underset{P^{\prime}\in\mathcal{P}(s,P)}{\text{ess}\sup}E^{P^{\prime}}%
[X\mid\mathcal{F}_{s}]\\
= & \underset{P^{\prime}\in\mathcal{P}(s,P)}{\text{ess}\sup}E^{P^{\prime}%
}[\underset{P^{\prime\prime}\in\mathcal{P}(t,P^{\prime})}{\text{ess}\sup
}E^{P^{\prime\prime}}[X\mid\mathcal{F}_{t}]\mid\mathcal{F}_{s}]
\end{array}
\label{exttc4}%
\end{equation}
The classical law of iterated expectations implies the inequality $\leq$ in
(\ref{exttc4}). Next prove the reverse inequality.

As in (\ref{exttc1}), there exists a sequence $P_{n}^{\prime\prime}%
\in\mathcal{P}(t,P^{\prime})$ such that $P^{\prime}$-$a.e.$%
\[
\underset{n\rightarrow\infty}{\lim}E^{P_{n}^{\prime\prime}}[X\mid
\mathcal{F}_{t}]\uparrow\underset{P^{\prime\prime}\in\mathcal{P}(t,P^{\prime
})}{\text{ess}\sup}E^{P^{\prime\prime}}[X\mid\mathcal{F}_{t}]
\]
Then
\[%
\begin{array}
[c]{rl}
& E^{P^{\prime}}[\underset{P^{\prime\prime}\in\mathcal{P}(t,P^{\prime
})}{\text{ess}\sup}E^{P^{\prime\prime}}[X\mid\mathcal{F}_{t}]\mid
\mathcal{F}_{s}]\\
= & \underset{n\rightarrow\infty}{\lim}E^{P_{n}^{\prime\prime}}[X\mid
\mathcal{F}_{s}]\\
\leq & \underset{\tilde{P}\in\mathcal{P}(s,P)}{\text{ess}\sup}E^{\tilde{P}%
}[X\mid\mathcal{F}_{s}]\;P\text{-}a.e.
\end{array}
\]
This proves (\ref{LIE}).

Proof properties (i)-(iv) is standard and is omitted. This completes the proof
of Theorem \ref{thm-conditioning}. \hfill$\blacksquare$

\section{Appendix: Proofs for Utility\label{app-utility}}

\noindent\textit{Proof of Theorem \ref{thm-utility}}: \textbf{Part (a).}
Consider the following backward stochastic differential equation under
$\hat{E}$:%
\[
V_{t}=\hat{E}[\xi+\int\nolimits_{t}^{T}f(c_{s},V_{s})ds\mid\mathcal{F}%
_{t}]\text{, }\;t\in\lbrack0,T]\text{,}%
\]
where $\xi\in\widehat{L^{2}}(\Omega_{T})$ is terminal utility and $c\in D$.
(Equation (\ref{Vt}) is the special case where $\xi=0$.) Given $V\in
M^{2}(0,T)$, let%
\[
\Lambda_{t}(V)\equiv\hat{E}[\xi+\int\nolimits_{t}^{T}f(c_{s},V_{s}%
)ds\mid\mathcal{F}_{t}],\;t\in\lbrack0,T]\text{.}%
\]

We need the following regularity property of $\Lambda.$

\begin{lemma}
$\Lambda$ is a mapping from $M^{2}(0,T)$ to $M^{2}(0,T)$.
\end{lemma}%

\textbf{Proof.}
By assumption (ii) there exists a positive constant $K$ such that
\[
\mid f(c_{s},V_{s})-f(c_{s},0)\mid\leq K\mid V_{s}\mid,\;s\in\lbrack0,T].
\]
Because both $V$ and $(f(c_{s},0))_{0\leq s\leq T}$ are in $M^{2}(0,T)$, we
have $(f(c_{s},V_{s}))_{0\leq s\leq T}\in M^{2}(0,T)$. Thus%

\[%
\begin{array}
[c]{rl}
& \hat{E}[\mid\xi+\int\nolimits_{t}^{T}f(c_{s},V_{s})ds\mid^{2}]\\
\leq & 2\hat{E}[\mid\xi\mid^{2}+(T-t)\int\nolimits_{t}^{T}\mid f(c_{s}%
,V_{s})\mid^{2}ds]\\
\leq & 2\hat{E}[\mid\xi\mid^{2}]+2(T-t)\hat{E}[\int\nolimits_{t}^{T}\mid
f(c_{s},V_{s})\mid^{2}ds]<\infty\text{,}%
\end{array}
\]
which means that $(\xi+\int\nolimits_{t}^{T}f(c_{s},V_{s})ds)\in
\widehat{L^{2}}(\Omega)$. Argue further that%

\[%
\begin{array}
[c]{rl}%
\hat{E}[\mid\Lambda_{t}(V)\mid^{2}] & =\hat{E}[(\hat{E}[\xi+\int%
\nolimits_{t}^{T}f(c_{s},V_{s})ds\mid\mathcal{F}_{t}])^{2}]\\
& \leq\hat{E}[\hat{E}[\mid\xi+\int\nolimits_{t}^{T}f(c_{s},V_{s})ds\mid
^{2}\mid\mathcal{F}_{t}]]\\
& =\hat{E}[\mid\xi+\int\nolimits_{t}^{T}f(c_{s},V_{s})ds\mid^{2}\mid
]<\infty\text{.}%
\end{array}
\]
Finally, $(\hat{E}%
{\displaystyle\int\nolimits_{0}^{T}}
\mid\Lambda_{t}\mid^{2}dt)^{\frac{1}{2}}\leq(%
{\displaystyle\int\nolimits_{0}^{T}}
\hat{E}[\mid\Lambda_{t}\mid^{2}]dt)^{\frac{1}{2}}<\infty$ and $\Lambda\in
M^{2}(0,T)$.\hfill$\blacksquare$

\bigskip

For any $V$ and $V^{\prime}\in M^{2}(0,T)$, by Theorem \ref{thm-conditioning}
the following approximation holds:%
\begin{equation}%
\begin{array}
[c]{rl}
& \mid\Lambda_{t}(V)-\Lambda_{t}(V^{\prime})\mid^{2}\\
= & (\hat{E}[\int\nolimits_{t}^{T}(f(c_{s},V_{s})-f(c_{s},V_{s}^{\prime
}))ds\mid\mathcal{F}_{t}])^{2}\\
\leq & \hat{E}[(\int\nolimits_{t}^{T}\mid f(c_{s},V_{s})-f(c_{s},V_{s}%
^{\prime})\mid ds)^{2}\mid\mathcal{F}_{t}]\\
\leq & (T-t)K^{2}\hat{E}[%
{\displaystyle\int\nolimits_{t}^{T}}
\mid V_{s}-V_{s}^{\prime}\mid^{2}ds\mid\mathcal{F}_{t}]\\
\leq & L\hat{E}[%
{\displaystyle\int\nolimits_{t}^{T}}
\mid V_{s}-V_{s}^{\prime}\mid^{2}ds\mid\mathcal{F}_{t}]\text{,}%
\end{array}
\label{L}%
\end{equation}
where $K$ is the Lipschitz constant for the aggregator and $L=TK^{2}$. (The
first inequality is due to the classical Jensen's inequality and
(\ref{Econditional}).) Then for each $r\in\lbrack0,T]$,%
\[%
\begin{array}
[c]{l}%
\;\hat{E}[%
{\displaystyle\int\nolimits_{r}^{T}}
\mid\Lambda_{t}(V)-\Lambda_{t}(V^{\prime})\mid^{2}dt]\\
\leq L\hat{E}[%
{\displaystyle\int\nolimits_{r}^{T}}
\hat{E}[%
{\displaystyle\int\nolimits_{t}^{T}}
\mid V_{s}-V_{s}^{\prime}\mid^{2}ds\mid\mathcal{F}_{t}]dt]\\
\leq L%
{\displaystyle\int\nolimits_{r}^{T}}
\hat{E}[%
{\displaystyle\int\nolimits_{t}^{T}}
\mid V_{s}-V_{s}^{\prime}\mid^{2}ds]dt\\
\leq L(T-r)\hat{E}[%
{\displaystyle\int\nolimits_{r}^{T}}
\mid V_{s}-V_{s}^{\prime}\mid^{2}ds].
\end{array}
\]
Set $\delta=\frac{1}{2L}$ and $r_{1}=\max\{T-\delta,0\}$. Then,
\[
\hat{E}[%
{\displaystyle\int\nolimits_{r_{1}}^{T}}
\mid\Lambda_{t}(V)-\Lambda_{t}(V^{\prime})\mid^{2}dt]\leq\frac{1}{2}\hat{E}[%
{\displaystyle\int\nolimits_{r_{1}}^{T}}
\mid V_{t}-V_{t}^{\prime}\mid^{2}dt],
\]
which implies that $\Lambda$ is a contraction mapping from $M^{2}(r_{1},T)$ to
$M^{2}(r_{1},T)$ and there exists a unique solution $(V_{t})\in M^{2}%
(r_{1},T)$ to the above BSDE. Because $\delta$ is independent of $t$, we can
work backwards in time and apply a similar argument at each step to prove that
there exists a unique solution $(V_{t})\in M^{2}(0,T)$.

\bigskip

\noindent\textbf{Part (b).} Uniqueness of the solution is due to the
contraction mapping property established in the proof of (a).

By Theorem \ref{thm-conditioning},%
\[%
\begin{array}
[c]{rl}
& -\,\widehat{E}\left[  -\int_{t}^{\tau}\,f(c_{s},V_{s})\,ds\,-\,V_{\tau}%
\mid\mathcal{F}_{t}\right] \\
= & -\,\widehat{E}\left[  -\int_{t}^{\tau}\,f(c_{s},V_{s})\,ds+\widehat{E}%
\left[  -\int_{\tau}^{T}\,f(c_{s},V_{s})\,ds\mid\mathcal{F}_{\tau}\right]
\mid\mathcal{F}_{t}\right] \\
= & -\widehat{E}[\widehat{E}\left[  -\int_{t}^{T}\,f(c_{s},V_{s}%
)\,ds\mid\,\mathcal{F}_{\tau}\right]  \mid\mathcal{F}_{t}]\\
= & -\widehat{E}[-\int_{t}^{T}\,f(c_{s},V_{s})\,ds\mid\mathcal{F}_{t}%
]~=~V_{t}%
\text{.\ \ \ \ \ \ \ \ \ \ \ \ \ \ \ \ \ \ \ \ \ \ \ \ \ \ \ \ \ \ \ \ \ \ \ \ }%
\blacksquare
\end{array}
\hfill
\]

\end{document}